\documentclass[12pt]{article}
\usepackage{amssymb,amsmath,epsfig}
\allowdisplaybreaks

\begin{document}

\title{\bf Extended Decoupled Anisotropic Solutions in $f(\mathcal{R},\mathcal{T},\mathcal{R}_{\gamma\chi}\mathcal{T}^{\gamma\chi})$ Gravity}
\author{M. Sharif \thanks{msharif.math@pu.edu.pk} and Tayyab Naseer \thanks{tayyabnaseer48@yahoo.com}\\
Department of Mathematics, University of the Punjab,\\
Quaid-i-Azam Campus, Lahore-54590, Pakistan.}

\date{}
\maketitle

\begin{abstract}
In this paper, we consider static spherical structure to develop
some anisotropic solutions by employing the extended gravitational
decoupling scheme in the background of
$f(\mathcal{R},\mathcal{T},\mathcal{R}_{\gamma\chi}\mathcal{T}^{\gamma\chi})$
gravity, where $\mathcal{R}$ and $\mathcal{T}$ indicate the Ricci
scalar and trace of the energy-momentum tensor, respectively. We
transform both radial as well as temporal metric functions and apply
them on the field equations that produce two different sets
corresponding to the decoupling parameter $\xi$. The first set is
associated with isotropic distribution, i.e., modified Krori-Barua
solution. The second set is influenced from anisotropic factor and
contains unknowns which are determined by taking some constraints.
The impact of decoupling parameter is then analyzed on the obtained
physical variables and anisotropy. We also investigate energy
conditions and some other parameters such as mass, compactness and
redshift graphically. It is found that our solution corresponding to
pressure-like constraint shows stable behavior throughout in this
gravity for the considered range of $\xi$.
\end{abstract}

\section{Introduction}

The structure of our cosmos is well-organized yet unfathomable, made
up of massive geometrical structures such as stars and galaxies as
well as other obscure components. The physical behavior of such
heavily bodies help to understand cosmic evolution. General
Relativity (GR) is the first relativistic theory which provides
basic understanding of both astrophysical as well as cosmological
phenomena. Numerous cosmological experiments performed on the far
away galaxies indicate that our universe is getting larger rapidly.
The repulsive form of energy, named as the dark energy, is
considered to execute such expansion. The modifications to GR have
therefore been found very important to disclose hidden features of
our universe. The $f(\mathcal{R})$ theory is the first ever
modification of GR in which the Ricci scalar $\mathcal{R}$ is
replaced with the functional $f(\mathcal{R})$ in the
Einstein-Hilbert action. Several researchers \cite{2}-\cite{4}
employed different techniques in this theory to study the physical
characteristics and stability of compact objects. Capozziello and
his associates \cite{8} utilized the Lan\'{e}-Emden equation to
analyze the stable configuration of different models in this theory.
The formation and stability of the astronomical objects have been
examined by various authors \cite{9}-\cite{9c}.

The matter-geometry coupling in $f(\mathcal{R})$ gravity was first
studied by Bertolami \emph{et al.} \cite{10} by considering the
Lagrangian in terms of $\mathcal{R}$ and $\mathcal{L}_{m}$. Such a
coupling in modified gravitational theories has led many astronomers
to focus on the accelerating expansion of our cosmos. Another
modified theory which involves matter-geometry interaction was
developed by Harko \emph{et al.} \cite{20} is the
$f(\mathcal{R},\mathcal{T})$ theory, $\mathcal{T}$ shows trace of
the energy-momentum tensor. This interaction supports the
energy-momentum tensor not to be conserved, which may result in the
accelerating cosmic expansion. Later, a more complicated theory,
named as $f(\mathcal{R},\mathcal{T},\mathcal{Q})$ gravity, was
established by Haghani \emph{et al.} \cite{22} to study the
influence of strong non-minimal interaction on interstellar
structures, where $\mathcal{Q}\equiv
\mathcal{R}_{\gamma\chi}\mathcal{T}^{\gamma\chi}$. They studied some
mathematical models and found stability criteria for them. They also
utilized Lagrange multiplier approach to obtain conserved equations
in this theory.

The construction of this gravity is based on the insertion of strong
non-minimal coupling between matter and geometry which is explained
by the factor $\mathcal{R}_{\gamma\chi}\mathcal{T}^{\gamma\chi}$.
The role of dark matter and dark energy, without resorting to exotic
matter distribution is explained through the modification in the
Einstein-Hilbert action. Some other theories like
$f(\mathcal{R},\mathcal{T})$ and $f(\mathcal{R},\mathcal{L}_m)$ also
engage such non-minimal coupling but their functionals cannot be
considered in the most general form to understand the effects of
coupling on massive bodies in some scenarios. It should be stressed
that the insertion of
$\mathcal{R}_{\gamma\chi}\mathcal{T}^{\gamma\chi}$ could explain
non-minimal interaction in the situations where
$f(\mathcal{R},\mathcal{T})$ theory fails to explain. For instance,
the gravitational equations of motion corresponding to the massive
particles cannot encompass the effects of non-minimal coupling in
$f(\mathcal{R},\mathcal{T})$ gravity in case of the trace-free
energy-momentum tensor $(\mathcal{T}=0)$, while
$f(\mathcal{R},\mathcal{T},\mathcal{Q})$ theory provides such
coupling even in this context. This theory was shown to be stable
against Dolgov-Kawasaki instability \cite{22}. The presence of an
extra force in this theory, which helps the test particles to move
in non-geodesic path, could explain the galactic rotation curves.

Sharif and Zubair considered FRW spacetime to study the laws of
black hole thermodynamics \cite{22a} for some particular models and
derived energy bounds \cite{22b} in this gravity. In
$f(\mathcal{R},\mathcal{T},\mathcal{Q})$ gravity, Odintsov and
S\'{a}ez-G\'{o}mez \cite{23} solved the complicated field equations
corresponding to different models through numerical technique and
highlighted some challenges related to the instability of matter
distribution. Ayuso \emph{et al.} \cite{24} assumed some appropriate
scalar/vector fields to obtain the stability conditions for this
theory and found that instability occurs necessarily for the vector
field. Baffou \emph{et al.} \cite{25} calculated gravitational field
equations for FRW geometry and solved them by introducing
perturbation functions to analyze their viability. Yousaf \emph{et
al.} \cite{26}-\cite{26e} calculated four structure scalars in this
theory which are found to be very useful to study the evolution of
static and non-static spacetimes. We used different approaches to
discuss charged/uncharged compact anisotropic objects and obtained
physically acceptable solutions to the field equations
\cite{27}-\cite{27c}.

The quest for accurate solutions of self-gravitating matter
distributions has always been a fascinating problem due to
non-linear nature of the field equations. Multiple approaches have
been used to find such solutions corresponding to the isotropic as
well as anisotropic configured celestial bodies. Sharif and Waseem
\cite{25a,25aa} investigated the viability and stability of
different compact objects in
$f(\mathcal{R},\mathcal{T},\mathcal{Q})$ gravity by using the
Krori-Barua solution. They found stable solution for the case of
anisotropic matter distribution, whereas isotropic configured stars
are shown to be unstable near the center. Maurya \emph{et al.}
\cite{25b} considered the model
$f(\mathcal{R},\mathcal{T})=\mathcal{R}+2\chi\mathcal{T}$ and
examined the physical feasibility of anisotropic structures through
the MIT bag model and Karmarkar condition. Shamir and Fayyaz
\cite{25c} discussed four anisotropic compact stars in
$f(\mathcal{R})$ theory by taking different models. A newly
established methodology named as the minimal geometric deformation
(MGD) via gravitational decoupling has shown its significant
consequences for the development of physically acceptable solutions.
This approach offers various noteworthy ingredients for accurate
solutions in the field of cosmology and astrophysics. Initially,
Ovalle developed \cite{29} this procedure to construct new
analytical solutions for spherically symmetric interstellar bodies
in the background of braneworld. Following the braneworld scenario,
Ovalle and Linares \cite{30} formulated exact solutions for the case
of isotropic spherical distribution which are compatible with the
Tolman IV solution. For spherical exterior spacetime, Casadio
\emph{et al.} \cite{31} constructed a new solution whose behavior is
singular at the Schwarzschild radius.

Ovalle \cite{32} employed the decoupling strategy to establish an
accurate anisotropic solution for spherical distribution. The
isotropic solution has been generalized by Ovalle \emph{et al.}
\cite{33} to new anisotropic solutions which are analyzed
graphically. Gabbanelli \emph{et al.} \cite{34} assumed
Duragpal-Fuloria isotropic solution and extended it to anisotropic
solution which is found to be physically feasible. Estrada and
Tello-Ortiz \cite{34a} formulated several anisotropic solutions from
Heintzmann isotropic ansatz and concluded that these solutions are
physically viable. Sharif and Sadiq \cite{35} formed two anisotropic
solutions by using the Krori-Barua spacetime in the presence of
electromagnetic field and explored the effect of decoupling
parameter on the matter variables as well as the energy conditions.
Sharif and his collaborators \cite{36}-\cite{36c} generalized these
solutions to modified theories like $f(\mathcal{G})$ and
$f(\mathcal{R})$, where $\mathcal{G}$ is the Gauss-Bonnet invariant.
Making use of the same strategy, Singh \emph{et al.} \cite{37}
calculated various anisotropic solutions by employing class-one
condition. Hensh and Stuchl\'{i}k \cite{37a} considered Tolman VII
solution and engaged an appropriate deformation to find different
viable anisotropic solutions.

Although the MGD (which transforms the radial metric potential only)
is a highly efficient approach to develop feasible solutions of the
field equations. However, this is possible only when the sources
under consideration do not exchange energy to each other. Ovalle
\cite{38} resolved this problem by presenting a new strategy in
which both (radial and temporal) metric functions have been
transformed. This strategy is known as the extended gravitational
decoupling (EGD), which works for all spacetime regions for any
choice of the matter configuration. For the case of
$2+1$-dimensional geometry, Contreras and Bargue\~{n}o \cite{39}
considered vacuum BTZ solution and extended it to the charged BTZ
solution by employing EGD scheme. This technique has been utilized
by Sharif and Ama-Tul-Mughani to construct anisotropic solutions by
extending the isotropic Tolman IV \cite{40} and Krori-Barua
\cite{41} solutions. Sharif and Saba \cite{41a} developed
anisotropic solutions in $f(\mathcal{G})$ gravity through this
technique. Sharif and Majid \cite{41b}-\cite{41d} used both minimal
and extended strategies to formulate various cosmological solutions
in Brans-Dicke scenario by considering different isotropic solutions
such as Krori-Barua and Tolman IV.

This paper explores the influence of
$f(\mathcal{R},\mathcal{T},\mathcal{Q})$ correction terms on
different anisotropic solutions corresponding to spherical geometry,
which are constructed via EGD scheme. The paper is structured in the
following way. We introduce some basic formulation of this modified
gravity in the next section. The EGD approach is discussed in
section 3, which helps to divide the field equations into two
sectors. In section 4, we develop two anisotropic solutions by
utilizing the Krori-Barua solution and examine their physical
feasibility. Lastly, we summarize our findings in section 5.

\section{The $f(\mathcal{R},\mathcal{T},\mathcal{R}_{\gamma\chi}\mathcal{T}^{\gamma\chi})$ Gravity}

The modification of the Einstein-Hilbert action involving an
additional source (with $\kappa=8\pi$) is given as \cite{23}
\begin{equation}\label{g1}
S_{f(\mathcal{R},\mathcal{T},\mathcal{R}_{\gamma\chi}\mathcal{T}^{\gamma\chi})}=\int
\left[\frac{f(\mathcal{R},\mathcal{T},\mathcal{R}_{\gamma\chi}\mathcal{T}^{\gamma\chi})}{16\pi}
+\mathcal{L}_{m}+\xi\mathcal{L}_{\Delta}\right]\sqrt{-g}d^{4}x,
\end{equation}
where $\mathcal{L}_{m}$ and $\mathcal{L}_{\Delta}$ refer to the
Lagrangian for matter distribution and new gravitational source,
respectively. Here, the matter Lagrangian is taken to be negative of
the energy density of fluid, i.e., $\mathcal{L}_{m}=-\mu$. The field
equations corresponding to the action \eqref{g1} become
\begin{equation}\label{g2}
\mathcal{G}_{\gamma\chi}=8\pi \mathcal{T}_{\gamma\chi}^{(tot)}.
\end{equation}
The quantity $\mathcal{G}_{\gamma\chi}$ indicates the Einstein
tensor which shows the geometry and
$\mathcal{T}_{\gamma\chi}^{(tot)}$ is the stress-energy tensor which
may further be expressed as
\begin{equation}\label{g3}
\mathcal{T}_{\gamma\chi}^{(tot)}=\mathcal{T}_{\gamma\chi}^{(eff)}+\xi
\Delta_{\gamma\chi}=\frac{1}{f_{\mathcal{R}}-\mathcal{L}_{m}f_{\mathcal{Q}}}\mathcal{T}_{\gamma\chi}
+\mathcal{T}_{\gamma\chi}^{(\mathcal{D})}+\xi\Delta_{\gamma\chi}.
\end{equation}
The self-gravitating star contains anisotropic effects due to the
presence of new source $\Delta_{\gamma\chi}$ through some scalar,
vector or tensor field whose influence on a system is governed by a
decoupling parameter $\xi$. Moreover, the energy-momentum tensor in
$f(\mathcal{R},\mathcal{T},\mathcal{Q})$ gravity which includes
usual as well as extra curvature terms can be seen as
\begin{eqnarray}
\nonumber
\mathcal{T}_{\gamma\chi}^{(\mathcal{D})}&=&-\frac{1}{\mathcal{L}_{m}f_{\mathcal{Q}}-f_{\mathcal{R}}}
\left[\left(f_{\mathcal{T}}+\frac{1}{2}\mathcal{R}f_{\mathcal{Q}}\right)\mathcal{T}_{\gamma\chi}
+\left\{\frac{\mathcal{R}}{2}(\frac{f}{\mathcal{R}}-f_{\mathcal{R}})-\mathcal{L}_{m}f_{\mathcal{T}}\right.\right.\\\nonumber
&-&\left.\frac{1}{2}\nabla_{\eta}\nabla_{\upsilon}(f_{\mathcal{Q}}\mathcal{T}^{\eta\upsilon})\right\}g_{\gamma\chi}
-(g_{\gamma\chi}\Box-\nabla_{\gamma}\nabla_{\chi})f_{\mathcal{R}}
-\frac{1}{2}\Box(f_{\mathcal{Q}}\mathcal{T}_{\gamma\chi})\\\label{g4}
&+&\nabla_{\eta}\nabla_{(\gamma}[\mathcal{T}_{\chi)}^{\eta}f_{\mathcal{Q}}]-2f_{\mathcal{Q}}\mathcal{R}_{\eta(\gamma}\mathcal{T}_{\chi)}^{\eta}
+2(f_{\mathcal{Q}}\mathcal{R}^{\eta\upsilon}+\left.f_{\mathcal{T}}g^{\eta\upsilon})\frac{\partial^2\mathcal{L}_{m}}
{\partial g^{\gamma\chi}\partial g^{\eta\upsilon}}\right],
\end{eqnarray}
where $f_{\mathcal{R}}=\frac{\partial
f(\mathcal{R},\mathcal{T},\mathcal{Q})}{\partial
\mathcal{R}},~f_{\mathcal{T}}=\frac{\partial
f(\mathcal{R},\mathcal{T},\mathcal{Q})}{\partial \mathcal{T}}$ and
$f_{\mathcal{Q}}=\frac{\partial
f(\mathcal{R},\mathcal{T},\mathcal{Q})}{\partial \mathcal{Q}}$.
Also, $\nabla_\chi$ is the covariant derivative and $\Box\equiv
\nabla_\chi\nabla^\chi$. As the matter Lagrangian does not depend on
the metric tensor, thus the last term on the right hand side of
equation \eqref{g4} becomes zero \cite{22}, i.e.,
$\frac{\partial^2\mathcal{L}_{m}} {\partial g^{\gamma\chi}\partial
g^{\eta\upsilon}}=0$. The energy-momentum tensor for perfect fluid
is
\begin{equation}\label{g5}
\mathcal{T}_{\gamma\chi}=(\mu+\mathcal{P}) \mathcal{V}_{\gamma}
\mathcal{V}_{\chi}+\mathcal{P}g_{\gamma\chi},
\end{equation}
where $\mathcal{V}_{\chi}$ is the four-velocity and $\mathcal{P}$
shows isotropic pressure. The trace of the field equations yields
\begin{align}\nonumber
&3\nabla^{\chi}\nabla_{\chi}
f_\mathcal{R}+\mathcal{R}\left(f_\mathcal{R}-\frac{\mathcal{T}}{2}f_\mathcal{Q}\right)-\mathcal{T}(f_\mathcal{T}+1)
+\nabla_\gamma\nabla_\chi(f_\mathcal{Q}\mathcal{T}^{\gamma\chi})+\frac{1}{2}
\nabla^{\chi}\nabla_{\chi}(f_\mathcal{Q}\mathcal{T})\\\nonumber
&-2f+2\mathcal{R}_{\gamma\chi}\mathcal{T}^{\gamma\chi}f_\mathcal{Q}+(\mathcal{R}f_\mathcal{Q}+4f_\mathcal{T})\mathcal{L}_m
-2g^{\eta\upsilon} \frac{\partial^2\mathcal{L}_m}{\partial
g^{\eta\upsilon}\partial
g^{\gamma\chi}}\left(f_\mathcal{T}g^{\gamma\chi}+f_\mathcal{Q}R^{\gamma\chi}\right)=0.
\end{align}
In view of $\mathcal{Q}=0$ in the above equation,
$f(\mathcal{R},\mathcal{T})$ theory can be obtained whereas one can
retrieve $f(\mathcal{R})$ gravity by assuming the vacuum case.

The geometry, we consider, contains interior and exterior regions
which are separated by the hypersurface $\Sigma$. The following
metric defines static spherically symmetric object
\begin{equation}\label{g6}
ds^2=-e^{\psi} dt^2+e^{\phi} dr^2+r^2d\theta^2+r^2\sin^2\theta
d\varphi^2,
\end{equation}
where $\psi=\psi(r)$ and $\phi=\phi(r)$. The four-velocity and
four-vector in the comoving frame take the form
\begin{equation}\label{g7}
\mathcal{V}^\chi=(e^{\frac{-\psi}{2}},0,0,0),\quad
\mathcal{W}^\chi=(0,e^{\frac{-\phi}{2}},0,0),
\end{equation}
with the relations $\mathcal{V}^\chi \mathcal{V}_{\chi}=-1$ and
$\mathcal{W}^\chi \mathcal{V}_{\chi}=0$. For self-gravitating
geometry \eqref{g6}, the components of the field equations are
\begin{align}\label{g8}
&e^{-\phi}\left(\frac{\phi'}{r}-\frac{1}{r^2}\right)
+\frac{1}{r^2}=8\pi\left(\tilde{\mu}-\mathcal{T}_{0}^{0(\mathcal{D})}-\xi
\Delta_{0}^{0}\right),\\\label{g9}
&e^{-\phi}\left(\frac{\psi'}{r}+\frac{1}{r^2}\right)
-\frac{1}{r^2}=8\pi\left(\tilde{\mathcal{P}}+\mathcal{T}_{1}^{1(\mathcal{D})}+\xi\Delta_{1}^{1}\right),
\\\label{g10}
&\frac{e^{-\phi}}{4}\left[2\psi''+\psi'^2-\psi'\phi'+\frac{2\psi'}{r}-\frac{2\phi'}{r}\right]
=8\pi\left(\tilde{\mathcal{P}}+\mathcal{T}_{2}^{2(\mathcal{D})}+\xi\Delta_{2}^{2}\right),
\end{align}
where $\tilde{\mu}=\frac{1}{(\mu
f_{\mathcal{Q}}+f_{\mathcal{R}})}\mu$ and
$\tilde{\mathcal{P}}=\frac{1}{(\mu
f_{\mathcal{Q}}+f_{\mathcal{R}})}\mathcal{P}$. The presence of
modified correction terms
$\mathcal{T}_{0}^{0(\mathcal{D})},~\mathcal{T}_{1}^{1(\mathcal{D})}$
and $\mathcal{T}_{2}^{2(\mathcal{D})}$ make the above set of
equations \eqref{g8}-\eqref{g10} highly complex. Their values are
presented in Appendix \textbf{A}. Here, prime means
$\frac{\partial}{\partial r}$. Due to the matter-geometry
interaction in this theory, the stress-energy tensor exhibits
non-zero divergence, (i.e., $\nabla_\gamma
\mathcal{T}^{\gamma\chi}\neq 0$). Thus this theory disobeys the
equivalence principle as opposed to GR and $f(\mathcal{R})$ theory,
and the presence of an extra force leads to the non-geodesic motion
of particles in the gravitational field. Therefore we have
\begin{align}\nonumber
\nabla^\gamma
\mathcal{T}_{\gamma\chi}&=\frac{2}{2f_\mathcal{T}+\mathcal{R}f_\mathcal{Q}+1}\left[\nabla_\gamma(f_\mathcal{Q}\mathcal{R}^{\eta\gamma}
\mathcal{T}_{\eta\chi})+\nabla_\chi(\mathcal{L}_mf_\mathcal{T})-\frac{1}{2}
(f_\mathcal{T}g_{\eta\upsilon}+f_\mathcal{Q}\mathcal{R}_{\eta\upsilon})\right.\\\label{g11}
&\times\left.\nabla_\chi\mathcal{T}^{\eta\upsilon}-\mathcal{G}_{\gamma\chi}\nabla^\gamma(f_\mathcal{Q}\mathcal{L}_m)
-\frac{1}{2}\big\{\nabla^{\gamma}(\mathcal{R}f_{\mathcal{Q}})+2\nabla^{\gamma}f_{\mathcal{T}}\big\}\mathcal{T}_{\gamma\chi}\right].
\end{align}
This leads to the hydrostatic equilibrium condition as
\begin{align}\label{g12}
&\frac{d\mathcal{P}}{dr}+\xi\frac{d\Delta_{1}^{1}}{dr}+\frac{\psi'}{2}\left(\mu+\mathcal{P}\right)-\frac{2\xi}{r}
\left(\Delta_{2}^{2}-\Delta_{1}^{1}\right)-\frac{\xi\psi'}{2}\left(\Delta_{0}^{0}-\Delta_{1}^{1}\right)=\Omega,
\end{align}
which can be referred as the generalization of
Tolman-Opphenheimer-Volkoff (TOV) equation and the term $\Omega$,
appears due to non-zero divergence in
$f(\mathcal{R},\mathcal{T},\mathcal{Q})$ gravity, is provided in
Appendix \textbf{A}. This equation has been viewed to be very
significant to study the systematic changes in self-gravitating
celestial objects.

We have a set of four differential equations \eqref{g8}-\eqref{g10}
and \eqref{g12} which are highly non-linear and contains seven
unknowns
$(\psi,\phi,\mu,\mathcal{P},\Delta_{0}^{0},\Delta_{1}^{1},\Delta_{2}^{2})$,
thus the system becomes indeterminate. We use standardized approach
\cite{33} to get a determinate system so that we can calculate
unknown quantities. The effective matter variables can be expressed
as
\begin{equation}\label{g13}
\hat{\mu}=\tilde{\mu}-\xi\Delta_{0}^{0},\quad
\hat{\mathcal{P}}_{r}=\tilde{\mathcal{P}}+\xi\Delta_{1}^{1}, \quad
\hat{\mathcal{P}}_{\bot}=\tilde{\mathcal{P}}+\xi\Delta_{2}^{2},
\end{equation}
which indicate that the source $\Delta_{\gamma}^{\chi}$ leads to
anisotropic self-gravitating object. The effective anisotropy is
defined as
\begin{equation}\label{g14}
\hat{\Pi}=\hat{\mathcal{P}}_{\bot}-\hat{\mathcal{P}}_{r}=\xi\left(\Delta_{2}^{2}-\Delta_{1}^{1}\right),
\end{equation}
which vanishes for $\xi=0$.

\section{Extended Gravitational Decoupling}

In the following, we work out the system \eqref{g8}-\eqref{g10} and
calculate unknowns with the help of a novel algorithm, known as
gravitational decoupling through EGD approach. This modifies the
field equations so that the additional source
$\Delta_{\gamma}^{\chi}$ provides transformed equations which may
induce anisotropy in the inner region of celestial structure. We
begin with fundamental ingredient of this approach, i.e., the
perfect fluid solution $(\omega,\rho,\mu,\mathcal{P})$ with the
metric
\begin{equation}\label{g15}
ds^2=-e^{\omega}dt^2+e^{\rho}dr^2+r^2d\theta^2+r^2\sin^2\theta
d\varphi^2,
\end{equation}
where $\omega=\omega(r)$ and $\rho=\rho(r)=1-\frac{2m(r)}{r}$,
$m(r)$ is the Misner-Sharp mass of the geometry \eqref{g15}.
Further, we enforce linear geometrical transformations on both
metric potentials to examine the effects of additional source
$\Delta_{\gamma}^{\chi}$ on the isotropic solution as
\begin{equation}\label{g16}
\omega\rightarrow\psi=\omega+\xi \mathfrak{l}, \quad
e^{-\rho}\rightarrow e^{-\phi}=e^{-\rho}+\xi \mathfrak{n},
\end{equation}
where $\mathfrak{l}=\mathfrak{l}(r)$ and
$\mathfrak{n}=\mathfrak{n}(r)$ correspond to temporal and radial
metric components, respectively, thus EGD technique assures the
translation of both components.

We apply the transformation \eqref{g16} on equations
\eqref{g8}-\eqref{g10} to solve the complex field equations and get
two different sectors in which the first one for $\xi=0$ is given as
\begin{align}\label{g18}
&8\pi\left(\tilde{\mu}-\mathcal{T}_{0}^{0(\mathcal{D})}\right)=\frac{1}{r^2}+e^{-\rho}\left(\frac{\rho'}{r}-\frac{1}{r^2}\right),\\\label{g19}
&8\pi\left(\tilde{\mathcal{P}}+\mathcal{T}_{1}^{1(\mathcal{D})}\right)=-\frac{1}{r^2}+e^{-\rho}\left(\frac{\omega'}{r}+\frac{1}{r^2}\right),\\\label{g20}
&8\pi\left(\tilde{\mathcal{P}}+\mathcal{T}_{2}^{2(\mathcal{D})}\right)=e^{-\rho}\left(\frac{\omega''}{2}+\frac{\omega'^2}{4}-\frac{\omega'\rho'}{4}
+\frac{\omega'}{2r}-\frac{\rho'}{2r}\right),
\end{align}
and the second sector which contains anisotropic source
$\Delta^{\chi}_{\gamma}$ becomes
\begin{align}\label{g21}
8\pi\Delta_{0}^{0}&=\frac{\mathfrak{n}'}{r}+\frac{\mathfrak{n}}{r^2},\\\label{g22}
8\pi\Delta_{1}^{1}&=\mathfrak{n}\left(\frac{\psi'}{r}+\frac{1}{r^2}\right)+\frac{e^{-\rho}\mathfrak{l}'}{r},\\\nonumber
8\pi\Delta_{2}^{2}&=\frac{\mathfrak{n}}{4}\left(2\psi''+\psi'^2+\frac{2\psi'}{r}\right)+\frac{e^{-\rho}}{4}\left(2\mathfrak{l}''
+\xi\mathfrak{l}'^2+\frac{2\mathfrak{l}'}{r}+2\omega'\mathfrak{l}'-\rho'\mathfrak{l}'\right)\\\label{g23}
&+\frac{\mathfrak{n}'}{4}\left(\psi' +\frac{2}{r}\right).
\end{align}
The set of equations \eqref{g21}-\eqref{g23} are different from the
field equations for perfect matter distribution by few terms. Thus,
these equations can be labeled as the standard field equations for
anisotropic spherical geometry defined as
\begin{equation}\label{g24}
\Delta^{*\chi}_{\gamma}=\Delta^{\chi}_{\gamma}-\frac{1}{r^2}\delta_{\gamma}^{0}\delta_{0}^{\chi}
-\left(\mathcal{A}_{1}+\frac{1}{r^2}\right)\delta_{\gamma}^{1}\delta_{1}^{\chi}-\mathcal{A}_{2}\delta_{\gamma}^{2}\delta_{2}^{\chi},
\end{equation}
with explicit notations
\begin{align}\label{g24a}
\Delta^{*0}_{0}&=\Delta^{0}_{0}-\frac{1}{r^2},\\\label{g24b}
\Delta^{*1}_{1}&=\Delta^{1}_{1}-\left(\mathcal{A}_{1}+\frac{1}{r^2}\right),\\\label{g24c}
\Delta^{*2}_{2}&=\Delta^{2}_{2}-\mathcal{A}_{2},
\end{align}
with
\begin{equation}\nonumber
\mathcal{A}_{1}=\frac{e^{-\rho}\mathfrak{l}'}{r}, \quad
\mathcal{A}_{2}=\frac{e^{-\rho}}{4}\left(2\mathfrak{l}''
+\xi\mathfrak{l}'^2+\frac{2\mathfrak{l}'}{r}+2\omega'\mathfrak{l}'-\rho'\mathfrak{l}'\right).
\end{equation}
Consequently, the EGD technique has split the indefinite system
\eqref{g8}-\eqref{g10} into two sectors, the first of them
\eqref{g18}-\eqref{g20} indicates the field equations for isotropic
matter $(\tilde{\mu},\tilde{\mathcal{P}},\psi,\phi$). The other set
\eqref{g21}-\eqref{g23} follows the anisotropic system \eqref{g24}
and encompasses five unknowns
($\mathfrak{l},\mathfrak{n},\Delta_{0}^{0},\Delta_{1}^{1},\Delta_{2}^{2}$).
Finally, the system \eqref{g8}-\eqref{g10} has been successfully
decoupled.

\section{Anisotropic Solutions}

Our objective is to develop anisotropic solutions by employing
different constraints for compact object through EGD approach. For
this purpose, the field equations \eqref{g18}-\eqref{g20} need
isotropic solution in $f(\mathcal{R},\mathcal{T},\mathcal{Q})$
scenario. We adopt the non-singular Krori-Barua isotropic solution
\cite{42} to continue our analysis. Thus the solution in this
gravity becomes
\begin{eqnarray}\label{g33}
e^{\psi}&=&e^{Xr^2+Y}, \\\label{g34}
e^{\phi}&=&e^{\rho}=e^{Zr^2},\\\label{g35}
\tilde{\mu}&=&-\frac{1}{8\pi r^2}\left[e^{-Zr^2}
\left(1-2Zr^2\right)-1\right]+\mathcal{T}^{0(\mathcal{D})}_{0},
\\\label{g36} \mathcal{\tilde{P}}&=&\frac{1}{8\pi
r^2}\left[e^{-Zr^2}\left(1+2Xr^2\right)-1\right]-\mathcal{T}^{1(\mathcal{D})}_{1}.
\end{eqnarray}
The above equations involve three unknown quantities $X,~Y$ and $Z$
which can be found through matching conditions. After matching the
inner and outer spacetimes smoothly at boundary $r=H$, the
continuity of metric coefficients produce these constants as
\begin{eqnarray}\label{g37}
X&=&\frac{\breve{M}}{H^3}\left(1-\frac{2\breve{M}}{H}\right)^{-1},
\\\label{g38} Y&=&\ln\left(\frac{H-2\breve{M}}
{H}\right)-\frac{\breve{M}}{H}\left(1-\frac{2\breve{M}}{H}\right)^{-1},\\\label{g38a}
Z&=&\frac{1}{H^2} \ln\left(\frac{H}{H-2\breve{M}}\right),
\end{eqnarray}
with compactness $\frac{2\breve{M}}{H}<\frac{8}{9}$ and $\breve{M}$
is the total mass at the boundary. These equations pledge the
continuity of isotropic solution \eqref{g33}-\eqref{g36} with the
exterior Schwarzschild at the boundary which may be modified in the
interior due to the insertion of new source $\Delta_{\gamma\chi}$.
The inner spherical solutions involving pressure anisotropy
($\xi\neq 0$) can be formulated by using the metric functions
presented in equations \eqref{g33} and \eqref{g34}. Equations
\eqref{g21}-\eqref{g23} establish an interesting relation between
two deformation functions ($\mathfrak{l},\mathfrak{n}$) and source
$\Delta_{\gamma\chi}$, the solution of which can be calculated by
applying some conditions. In the following, we present some
constraints to obtain anisotropic solutions and check their
feasibility through graphical analysis.

\subsection{Solution-I}

Here, we calculate anisotropic solution by using a linear equation
of state given as
\begin{equation}\label{g38b}
\Delta^{0}_{0}=\alpha\Delta^{1}_{1}+\beta\Delta^{2}_{2},
\end{equation}
along with a restriction on $\Delta_{1}^{1}$ to figure out
$\mathfrak{l},~\mathfrak{n}$ and $\Delta_{\gamma}^{\chi}$. For our
ease, we choose $\alpha=1$ and $\beta=0$ so the relation
\eqref{g38b} becomes $\Delta^{0}_{0}=\Delta^{1}_{1}$. The interior
isotropic configuration is found to be compatible with exterior
Schwarzschild as long as
$\tilde{\mathcal{P}}(H)+\mathcal{T}_{1}^{1(\mathcal{D})}(H)\sim -\xi
\left(\Delta_{1}^{1}(H)\right)_{-}$. The straightforward choice
which fulfills this criteria is \cite{33}
\begin{equation}\label{g39}
-\tilde{\mathcal{P}}-\mathcal{T}_{1}^{1(\mathcal{D})}=\Delta_{1}^{1}.
\end{equation}
Making use of the field equations \eqref{g19},~\eqref{g21} and
\eqref{g22} with metric components \eqref{g33} and \eqref{g34}, we
obtain deformation functions as
\begin{eqnarray}\nonumber
\mathfrak{l}&=&\int \frac{1}{r\varpi}\left[\sqrt{\pi}e^{Z
r^2}(X+Z)\left(2Xr^2+1\right)\text{erf}\left(\sqrt{Z}r\right)-2
\sqrt{Z}r\left\{X\left(2Xr^2\right.\right.\right.\\\label{g39a}
&+&\left.\left.\left.1\right)+Z\left(2 X r^2\left(e^{Z
r^2}+1\right)+1\right)\right\}\right]dr,\\\label{g39b}
\mathfrak{n}&=&1-\frac{\sqrt{\pi}\left(X+Z\right)
\text{erf}\left(\sqrt{Z} r\right)}{2Z^{3/2}r}+\frac{Xe^{-Z r^2}}{Z},
\end{eqnarray}
where
\begin{equation}\nonumber
\varpi=2\sqrt{Z}r\left(Z\xi e^{Zr^2}+Z+X\xi\right)-\sqrt{\pi
}\xi\left(X+Z\right)e^{Zr^2}\text{erf}\left(\sqrt{Z} r\right).
\end{equation}
Thus the temporal and radial components take the form
\begin{eqnarray}\nonumber
\psi&=&Xr^2+Y+\xi \int\frac{1}{r\varpi}\left[\sqrt{\pi}e^{Z
r^2}(X+Z)\left(2Xr^2+1\right)\text{erf}\left(\sqrt{Z}r\right)\right.\\\label{g40}
&-&\left.2 \sqrt{Z}r\left\{X\left(2Xr^2+1\right)+Z\left(2 X
r^2\left(e^{Z r^2}+1\right)+1\right)\right\}\right]dr,\\\label{g40a}
e^{-\phi}&=&e^{-Zr^2}+\xi\left(1-\frac{\sqrt{\pi}\left(X+Z\right)
\text{erf}\left(\sqrt{Z} r\right)}{2Z^{3/2}r}+\frac{Xe^{-Z
r^2}}{Z}\right),
\end{eqnarray}
which will become standard Krori-Barua solution for perfect sphere
$(\xi=0)$.

To analyze the influence of anisotropy on triplet $(X,Y,Z)$, we use
the matching conditions at the boundary. The first fundamental form
produces the following expressions
\begin{eqnarray}\nonumber
\ln\left(1-\frac{2\breve{M}}{H}\right)&=&XH^2+Y+\xi
\left[\int\frac{1}{r\varpi}\left[\sqrt{\pi}e^{Z
r^2}(X+Z)\left(2Xr^2+1\right)\right.\right.\\\nonumber
&\times&\text{erf}\left(\sqrt{Z}r\right)-2 \sqrt{Z}r\left\{Z\left(2
X r^2\left(e^{Z r^2}+1\right)+1\right)\right.\\\label{g41}
&+&\left.\left.\left.X\left(2Xr^2+1\right)\right\}\right]dr\right]_{r=H},
\end{eqnarray}
and
\begin{eqnarray}\label{g42}
1-\frac{2\breve{M}}{H}=e^{-ZH^2}+\xi\left(1-\frac{\sqrt{\pi}\left(X+Z\right)
\text{erf}\left(\sqrt{Z}H\right)}{2Z^{3/2}H}+\frac{Xe^{-Z
H^2}}{Z}\right).
\end{eqnarray}
Similarly, the second fundamental form
$(\tilde{\mathcal{P}}(H)+\mathcal{T}_{1}^{1(\mathcal{D})}(H)+\xi
\left(\Delta_{1}^{1}(H)\right)_{-}=0)$ gives
\begin{equation}\label{g43}
\tilde{\mathcal{P}}(H)+\mathcal{T}_{1}^{1(\mathcal{D})}(H)=0
\quad\Rightarrow\quad Z=\frac{\ln\left(1+2XH^2\right)}{H^2},
\end{equation}
which interlinks the constants $X$ and $Z$. Equations
\eqref{g41}-\eqref{g43} offer suitable conditions which are very
useful for smooth matching of inner and outer spacetimes. Moreover,
after using the constraints \eqref{g38b} and \eqref{g39}, the first
anisotropic solution
$(\hat{\mu},\hat{\mathcal{P}}_{r},\hat{\mathcal{P}}_{\bot})$ and
anisotropic factor ($\hat{\Pi}$) become
\begin{eqnarray}\label{g46}
\hat{\mu}&=&\frac{1}{8\pi r^2}\left[e^{-Zr^2}\left(2Zr^2-1+2\xi
Xr^2+\xi\right)+1-\xi+8\pi
r^2\mathcal{T}_{0}^{0(\mathcal{D})}\right],\\\label{g47}
\hat{\mathcal{P}}_{r}&=&\frac{1}{8\pi r^2}\left[\left(1-\xi\right)
\left\{e^{-Zr^2}\left(2Xr^2+1\right)-1\right\}-8\pi
r^2\mathcal{T}_{1}^{1(\mathcal{D})}\right], \\\nonumber
\hat{\mathcal{P}}_{\bot}&=&\frac{1}{8\pi}\left[Xe^{-Zr^2}\left(X
r^2+2\right)-XZr^2e^{-Zr^2}-Ze^{-Zr^2}+\frac{\xi}{4Z^{3/2}
r^3}\left\{\left(Xr^2\right.\right.\right.\\\nonumber
&+&\left.\left.1\right)e^{-Zr^2}\left(\sqrt{\pi}(X+Z)e^{Zr^2}
\text{erf}\left(\sqrt{Z}r\right)-2\sqrt{Z}r\left(2XZr^2+Z+X\right)\right)\right\}\\\nonumber
&+&\xi X\left(Xr^2+2\right)\left(-\frac{\sqrt{\pi}
(X+Z)\text{erf}\left(\sqrt{Z}r\right)}{2Z^{3/2}r}+\frac{Xe^{-Z
r^2}}{Z}+1\right)+\frac{\xi}{4\varpi}\\\nonumber
&\times&\left\{2Ze^{-Zr^2}\left(\sqrt{\pi}
(X+Z)e^{Zr^2}\left(2Xr^2+1\right)\text{erf}\left(\sqrt{Z}r\right)+X
\left(2Xr^2+1\right)\right.\right.\\\nonumber &-&\left.\left.2
\sqrt{Z}r\left(Z\left(2Xr^2\left(e^{Z
r^2}+1\right)+1\right)\right)\right)\right\}+\frac{\xi
e^{-Zr^2}}{4\varpi^2}\left\{-4\sqrt{\pi}\sqrt{Z}\right.\\\nonumber
&\times&r(X+Z)
e^{Zr^2}\text{erf}\left(\sqrt{Z}r\right)\left(2Z^2r^2(\xi -1)
\left(2Xr^2+1\right)-Z\left(4X^2r^4\right.\right.\\\nonumber
&\times&\left.\left(\xi e^{Zr^2}+1\right)+4Xr^2\left(2\xi
e^{Zr^2}+\xi+1\right)-\xi e^{Zr^2}-1\right)\\\nonumber &+&\left.X\xi
\left(-4X^2r^4-8Xr^2+1\right)\right)+4Zr^2 \left(4Z^3r^2e^{Zr^2}
\left(2Xr^2 (\xi-1)+\xi\right)\right.\\\nonumber
&-&Z^2\left(4X^2r^4\left(e^{Zr^2}+1\right)\left(\xi
\left(e^{Zr^2}-1\right)+2\right)+4Xr^2\left(2\xi
e^{2Zr^2}+2\right)\right.\\\nonumber &-&\left.2\xi
e^{Zr^2}+\xi-2\right)-2XZ\left(4X^2r^4 \left(\xi
e^{Zr^2}+1\right)-\xi e^{Zr^2}-1\right.\\\nonumber
&+&\left.\left.4Xr^2\left(2\xi e^{Zr^2}+\xi
+1\right)\right)+X^2\xi\left(-4X^2r^4-8Xr^2+1\right)\right)-\pi
\xi\\\label{g48}
&\times&\left.\left.(X+Z)^2e^{2Zr^2}\left(4X^2r^4+8Xr^2-1\right)\text{erf}\left(\sqrt{Z}r\right)^2\right\}-8\pi
\mathcal{T}_{1}^{1(\mathcal{D})}\right],\\\nonumber
\hat{\Pi}&=&\frac{1}{8\pi}\left[Xe^{-Zr^2}\left(X
r^2+2\right)-XZr^2e^{-Zr^2}-Ze^{-Zr^2}+\frac{\xi}{4Z^{3/2}
r^3}\left\{\left(Xr^2\right.\right.\right.\\\nonumber
&+&\left.\left.1\right)e^{-Zr^2}\left(\sqrt{\pi}(X+Z)e^{Zr^2}
\text{erf}\left(\sqrt{Z}r\right)-2\sqrt{Z}r\left(2XZr^2+Z+X\right)\right)\right\}\\\nonumber
&+&\xi X\left(Xr^2+2\right)\left(-\frac{\sqrt{\pi}
(X+Z)\text{erf}\left(\sqrt{Z}r\right)}{2Z^{3/2}r}+\frac{Xe^{-Z
r^2}}{Z}+1\right)+\frac{\xi}{4\varpi}\\\nonumber
&\times&\left\{2Ze^{-Zr^2}\left(\sqrt{\pi}
(X+Z)e^{Zr^2}\left(2Xr^2+1\right)\text{erf}\left(\sqrt{Z}r\right)+X
\left(2Xr^2+1\right)\right.\right.\\\nonumber &-&\left.\left.2
\sqrt{Z}r\left(Z\left(2Xr^2\left(e^{Z
r^2}+1\right)+1\right)\right)\right)\right\}+\frac{\xi
e^{-Zr^2}}{4\varpi^2}\left\{-4\sqrt{\pi}\sqrt{Z}\right.\\\nonumber
&\times&r(X+Z)
e^{Zr^2}\text{erf}\left(\sqrt{Z}r\right)\left(2Z^2r^2(\xi -1)
\left(2Xr^2+1\right)-Z\left(4X^2r^4\right.\right.\\\nonumber
&\times&\left.\left(\xi e^{Zr^2}+1\right)+4Xr^2\left(2\xi
e^{Zr^2}+\xi+1\right)-\xi e^{Zr^2}-1\right)+X\xi\\\nonumber
&\times&\left. \left(-4X^2r^4-8Xr^2+1\right)\right)+4Zr^2
\left(4Z^3r^2e^{Zr^2} \left(2Xr^2
(\xi-1)+\xi\right)\right.\\\nonumber
&-&Z^2\left(4X^2r^4\left(e^{Zr^2}+1\right)\left(\xi
\left(e^{Zr^2}-1\right)+2\right)+4Xr^2\left(2\xi
e^{2Zr^2}+2\right)\right.\\\nonumber &-&\left.2\xi
e^{Zr^2}+\xi-2\right)-2XZ\left(4X^2r^4 \left(\xi
e^{Zr^2}+1\right)-\xi e^{Zr^2}-1+4Xr^2\right.\\\nonumber
&\times&\left.\left.\left(2\xi e^{Zr^2}+\xi
+1\right)\right)+X^2\xi\left(-4X^2r^4-8Xr^2+1\right)\right)-\pi \xi
e^{2Zr^2}\\\nonumber
&\times&\left.(X+Z)^2\left(4X^2r^4+8Xr^2-1\right)\text{erf}\left(\sqrt{Z}r\right)^2\right\}-\left(1-\xi\right)
\left\{\frac{e^{-Zr^2}}{r^2}\right.\\\label{g49}
&\times&\left.\left.\left(2Xr^2+1\right)-\frac{1}{r^2}\right\}\right].
\end{eqnarray}

\subsection{Solution-II}

We consider density-like constraint to get another anisotropic
solution for spherical configuration \eqref{g6} which is taken as
\begin{equation}\label{g51}
\tilde{\mu}-\mathcal{T}_{0}^{0(\mathcal{D})}=\Delta_{0}^{0}.
\end{equation}
The deformation functions can be found by using equations
\eqref{g18},~\eqref{g21} and \eqref{g22} along with constraints
\eqref{g38b} and \eqref{g51} as
\begin{eqnarray}\label{g52}
\mathfrak{l}&=&\int\frac{2r\left(-Xe^{Zr^2}+X+Z\right)}{\xi\left(e^{Zr^2}-1\right)+1}dr,\\\label{g53}
\mathfrak{n}&=&1-e^{Zr^2}.
\end{eqnarray}
The matching conditions for this solution are expressed as
\begin{align}\label{g55}
\ln\left(1-\frac{2\breve{M}}{H}\right)&=XH^2+Y+\xi\left[\int\frac{2r\left(-Xe^{Zr^2}
+X+Z\right)}{\xi\left(e^{Zr^2}-1\right)+1}dr\right]_{r=H},\\\label{g56}
Z&=-\frac{1}{H^2}\ln\left[1-\frac{1}{1-\xi}\left(\frac{2\breve{M}}{H}\right)\right].
\end{align}
Finally, we have expressions of the corresponding anisotropic
solution as
\begin{eqnarray}\label{g57}
\hat{\mu}&=&\frac{1}{8\pi
r^2}\left[\left(1-\xi\right)\left\{e^{-Zr^2}\left(2Zr^2-1\right)+1\right\}+8\pi
r^2\mathcal{T}_{0}^{0(\mathcal{D})}\right],\\\label{g58}
\hat{\mathcal{P}}_{r}&=&\frac{1}{8\pi
r^2}\left[e^{-Zr^2}\left(2Xr^2+1+2\xi Zr^2-\xi\right)-1+\xi-8\pi
r^2\mathcal{T}_{1}^{1(\mathcal{D})}\right],\\\nonumber
\hat{\mathcal{P}}_{\bot}&=&\frac{e^{-Zr^2}}{8\pi\left(\xi\left(e^{Zr^2}-1\right)+1\right)^2}\left[-Z^2r^2
\xi\left(\xi
e^{Zr^2}-1\right)+Z\xi\left(5-2e^{Zr^2}\right)-Z\right.\\\nonumber
&+&Z\xi^3\left(e^{Zr^2}-1\right)^2+Z\xi
^2\left(6e^{Zr^2}-e^{2Zr^2}-5\right)+XZr^2\left\{\left(e^{Zr^2}-1\right)^2\right.\\\nonumber
&\times&\left.\xi^3-2\xi^2\left(-3e^{Zr^2}+e^{2Zr^2}+2\right)+\chi
\left(6-5e^{Zr^2}\right)-1\right\}+2X\\\nonumber
&+&2X\xi^3\left(e^{Zr^2}-1\right)^3+4X\xi^2\left(e^{Zr^2}-1\right)^2+4X
\xi\left(e^{Zr^2}-1\right)+X^2r^2\\\nonumber
&+&\left.\left\{\xi^3\left(e^{Zr^2}-1\right)^3+2\xi^2\left(e^{Zr^2}-1\right)^2+\xi\left(e^{Zr^2}-1\right)+1\right\}\right]\\\label{g59}
&-&\mathcal{T}_{1}^{1(\mathcal{D})},
\end{eqnarray}
and the anisotropy becomes
\begin{eqnarray}\nonumber
\hat{\Pi}&=&\frac{1}{8\pi}\left[\frac{e^{-Zr^2}}{\left(\xi\left(e^{Zr^2}-1\right)+1\right)^2}\left\{-Z^2r^2
\xi\left(\xi
e^{Zr^2}-1\right)+Z\xi\left(5-2e^{Zr^2}\right)-Z\right.\right.\\\nonumber
&+&Z\xi^3\left(e^{Zr^2}-1\right)^2+Z\xi
^2\left(6e^{Zr^2}-e^{2Zr^2}-5\right)+XZr^2\left(\left(e^{Zr^2}-1\right)^2\right.\\\nonumber
&\times&\left.\xi^3-2\xi^2\left(-3e^{Zr^2}+e^{2Zr^2}+2\right)+\chi
\left(6-5e^{Zr^2}\right)-1\right)+2X\\\nonumber
&+&2X\xi^3\left(e^{Zr^2}-1\right)^3+4X\xi^2\left(e^{Zr^2}-1\right)^2+4X
\xi\left(e^{Zr^2}-1\right)+X^2r^2\\\nonumber
&+&\left.\left(\xi^3\left(e^{Zr^2}-1\right)^3+2\xi^2\left(e^{Zr^2}-1\right)^2+\xi\left(e^{Zr^2}-1\right)+1\right)\right\}
-\frac{e^{-Zr^2}}{r^2}\\\label{g60} &\times&\left.\left(2Xr^2+1+2\xi
Zr^2-\xi\right)+1-\xi\right].
\end{eqnarray}

\subsection{Physical Interpretation of the Obtained Solutions}

For a self-gravitating spherical object, the mass can be expressed
as
\begin{equation}\label{g63}
m(r)=4\pi\int_{0}^{H}r^2 \hat{\mu}dr.
\end{equation}
The numerical solution of the above equation gives mass of the
geometry \eqref{g6}, where we have used an initial condition
$m(0)=0$. The ratio of mass and radius of a celestial object is
known as compactness $(\zeta(r))$ which is considered as a
significant property of a self-gravitating structure. Buchdahl
\cite{42a} developed an upper limit of $\zeta(r)$ by employing the
continuity of fundamental forms of junction conditions between inner
and outer spacetimes at the hypersurface. It is found that this
limit must be less than $\frac{4}{9}$ in stable region of stellar
configuration, where $m(r)=\frac{H}{2}\left(1-e^{-\phi}\right)$. The
increment in the wavelength of electromagnetic diffusion (occurs in
celestial body due to its robust gravitational pull) can be
calculated by a redshift parameter characterized as
$D(r)=\frac{1}{\sqrt{1-2\zeta}}-1$. Buchdahl hampered its value at
the boundary of the star as $D(r)<2$ for stable configuration in the
case of perfect fluid, while its upper limit has been found to be
5.211 for anisotropic structures \cite{42b}.

Another phenomenon which plays a major role in self-gravitating
geometry is the energy conditions. The fulfillment of such
constraints assure the existence of normal matter as well as
viability of the solutions. These conditions are followed by the
parameters which govern the interior stellar configuration involving
ordinary matter. These bounds have been classified in four different
categories, namely null, weak, strong and dominant energy conditions
which become in $f(\mathcal{R},\mathcal{T},\mathcal{Q})$ theory as
\begin{eqnarray}\nonumber
&&\hat{\mu} \geq 0, \quad \hat{\mu}+\hat{\mathcal{P}}_{r} \geq
0,\\\nonumber &&\hat{\mu}+\hat{\mathcal{P}}_{\bot} \geq 0, \quad
\hat{\mu}-\hat{\mathcal{P}}_{r} \geq 0,\\\label{g50}
&&\hat{\mu}-\hat{\mathcal{P}}_{\bot} \geq 0, \quad
\hat{\mu}+\hat{\mathcal{P}}_{r}+2\hat{\mathcal{P}}_{\bot} \geq 0.
\end{eqnarray}
To study the feasibility of a star in astrophysics, the stability
plays a significant role. Here, we employ two different criteria to
analyze the stable regions of interior geometry for the obtained
solutions. We use causality condition which requires the squared
sound speed to be within $(0,1)$, i.e., $0 < v_{s}^{2} < 1$. For
anisotropic fluid, the difference between squared sound speeds in
both tangential
$(v_{s\bot}^{2}=\frac{d\hat{\mathcal{P}}_{\bot}}{d\hat{\mu}})$ and
radial directions
$(v_{sr}^{2}=\frac{d\hat{\mathcal{P}}_{r}}{d\hat{\mu}})$ is used to
check the stability of compact stars as $\mid
v_{s\bot}^{2}-v_{sr}^{2} \mid < 1$. The other influential criterion
to examine stability is the adiabatic index $(\Upsilon)$. The stable
region of the astronomical object must have its value greater than
$\frac{4}{3}$ \cite{42c}. Here, $\Upsilon$ has the form
\begin{equation}\label{g62}
\hat{\Upsilon}=\frac{\hat{\mu}+\hat{\mathcal{P}}_{r}}{\hat{\mathcal{P}}_{r}}
\left(\frac{d\hat{\mathcal{P}}_{r}}{d\hat{\mu}}\right).
\end{equation}
\begin{figure}\center
\epsfig{file=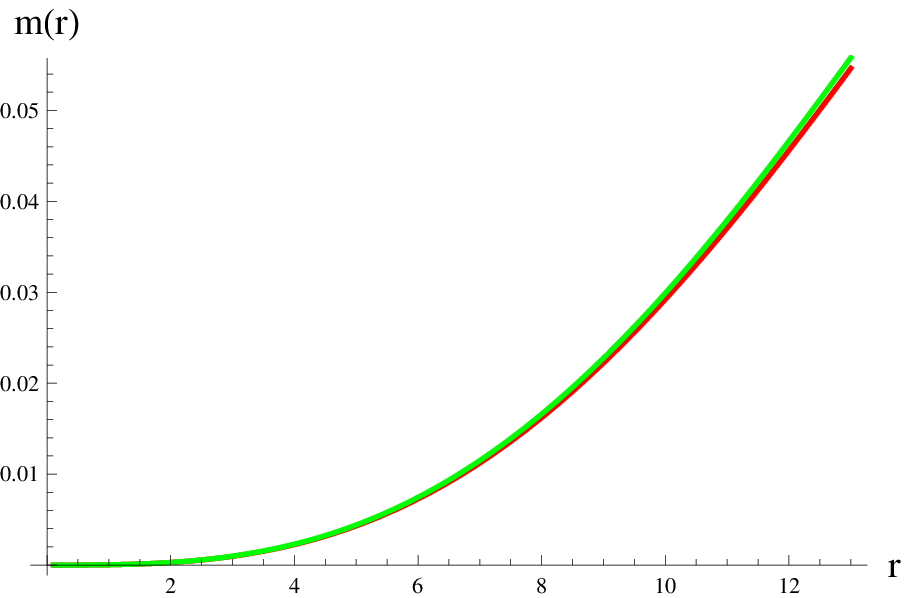,width=0.4\linewidth}\epsfig{file=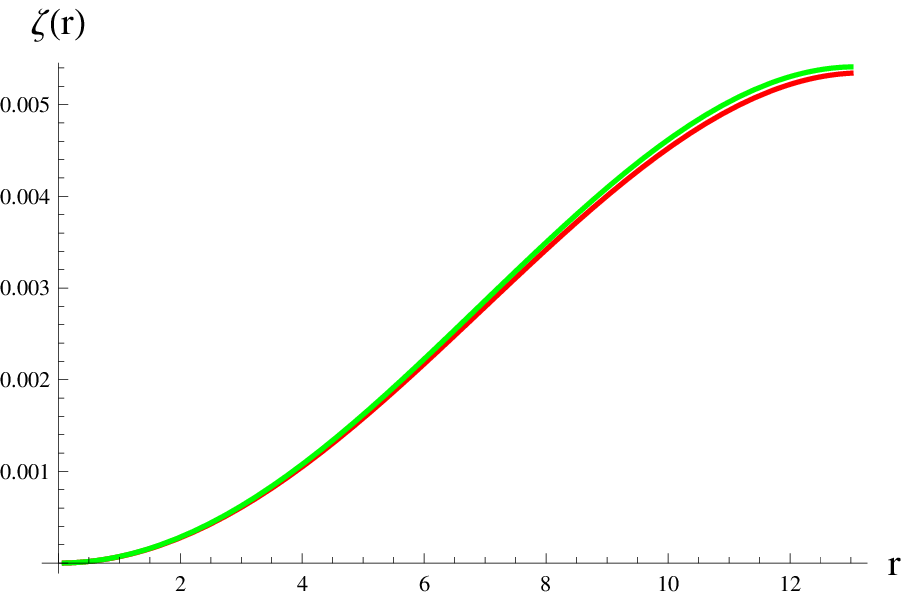,width=0.4\linewidth}
\epsfig{file=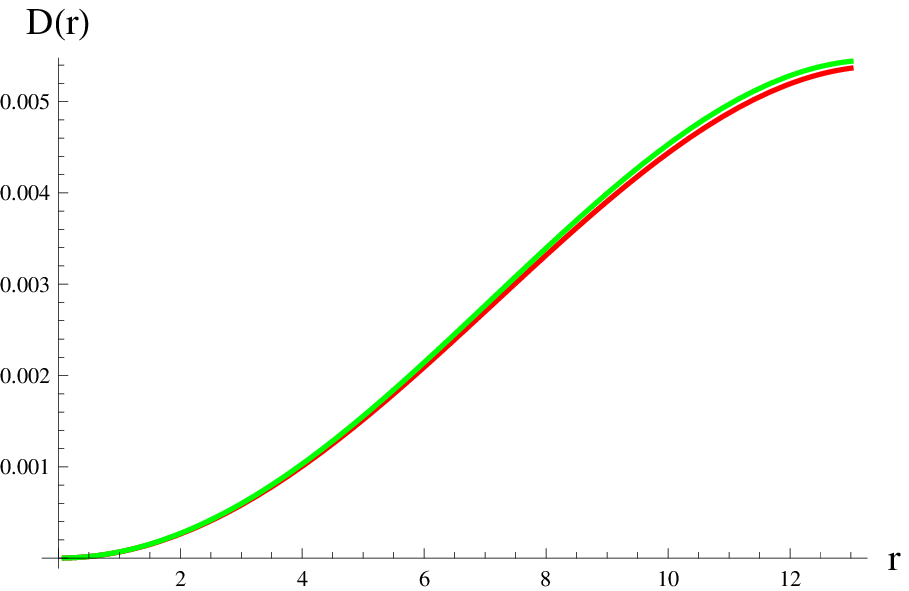,width=0.4\linewidth} \caption{Plots of
mass, compactness and redshift parameters corresponding to $\xi=0.2$
(red) and $\xi=0.8$ (green) for Solution-I.}
\end{figure}

As this theory involves complicated functional form, thus for our
ease, we adopt a linear model \cite{22} engaging an arbitrary
constant $\varrho$ to investigate the physical feasibility of the
obtained solutions as
\begin{equation}\label{g61}
f(\mathcal{R},\mathcal{T},\mathcal{R}_{\gamma\chi}\mathcal{T}^{\gamma\chi})=\mathcal{R}+\varrho
\mathcal{R}_{\gamma\chi}\mathcal{T}^{\gamma\chi}.
\end{equation}
We set $\varrho=-0.1$ and fix the constant $Z$ given in equation
\eqref{g43} to analyze Solution-I physically. Equations \eqref{g37}
and \eqref{g38} present the values of parameters $X$ and $Y$. As the
above model encompasses the contraction of Ricci tensor with the
energy-momentum tensor, thus the effects of non-minimal
matter-geometry interaction can still be entailed on the massive
test particles. The mass of anisotropic sphere \eqref{g6} is
analyzed in Figure \textbf{1} (upper left) for two values of the
decoupling parameter, $\xi=0.1$ and $0.9$. We find that the mass
increases slowly with rise in $\xi$. Figure \textbf{1} (right and
lower) displays the compactness and redshift parameters of the
considered structure whose ranges are confirmed to be within their
respective bounds.
\begin{figure}\center
\epsfig{file=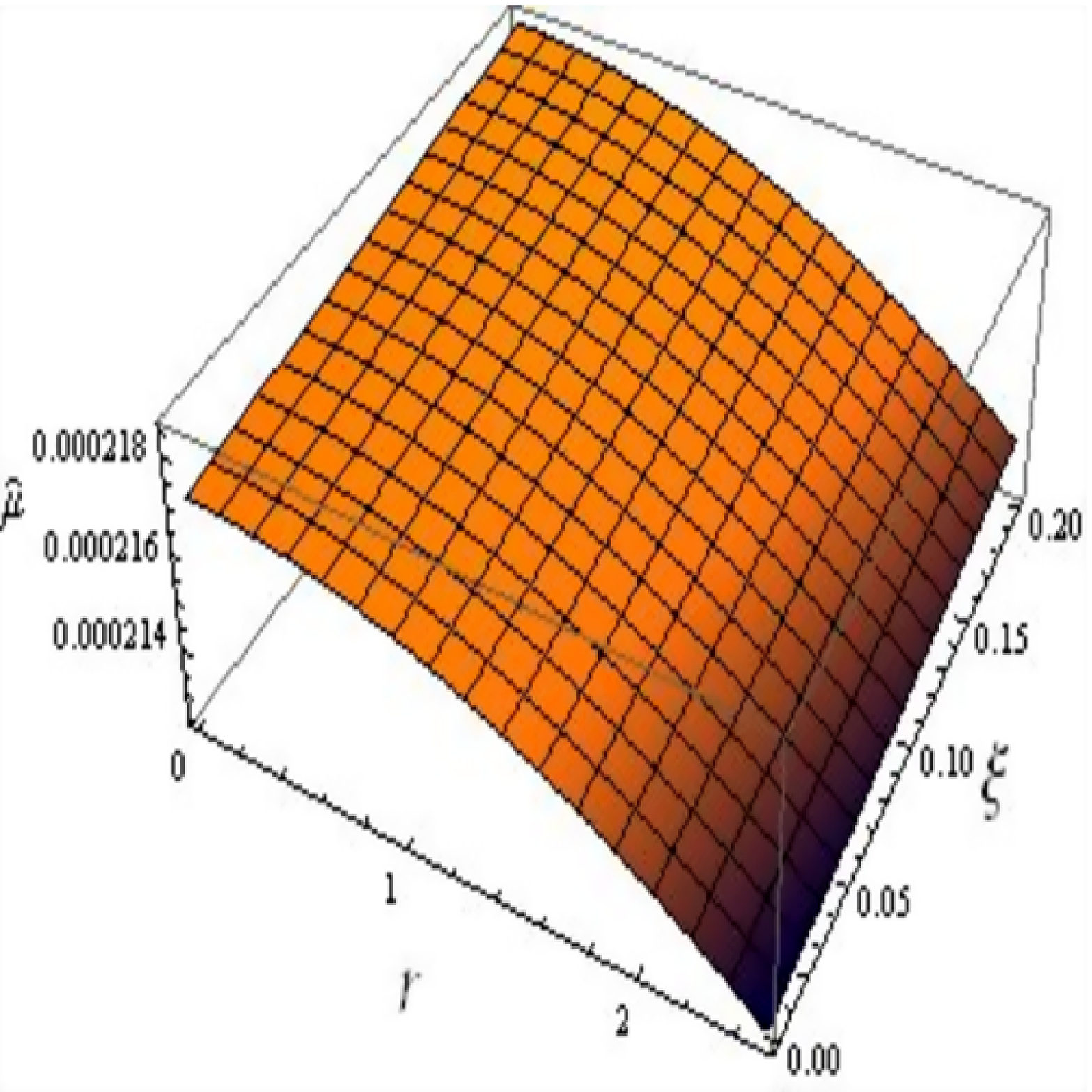,width=0.4\linewidth}\epsfig{file=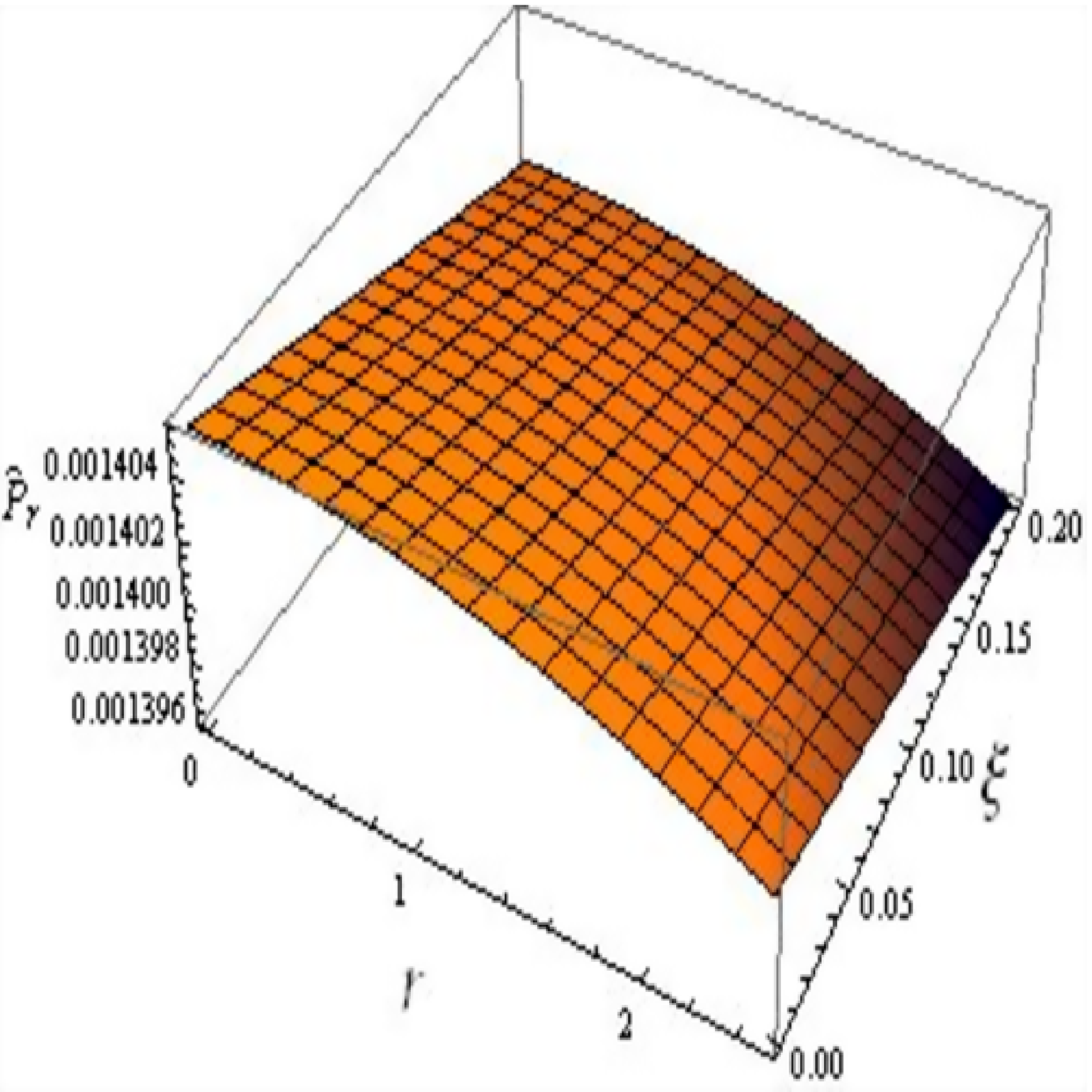,width=0.4\linewidth}
\epsfig{file=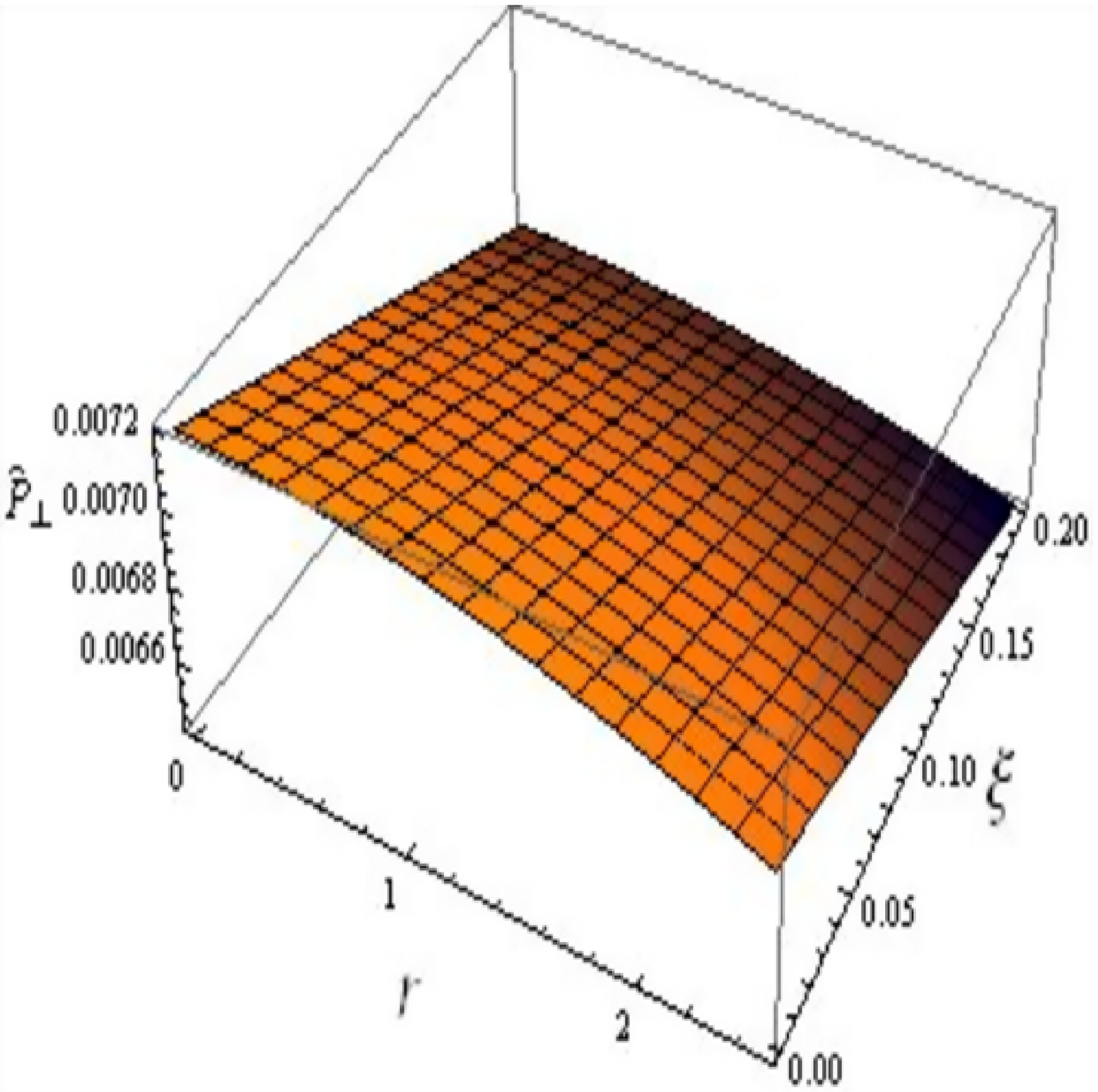,width=0.4\linewidth}\epsfig{file=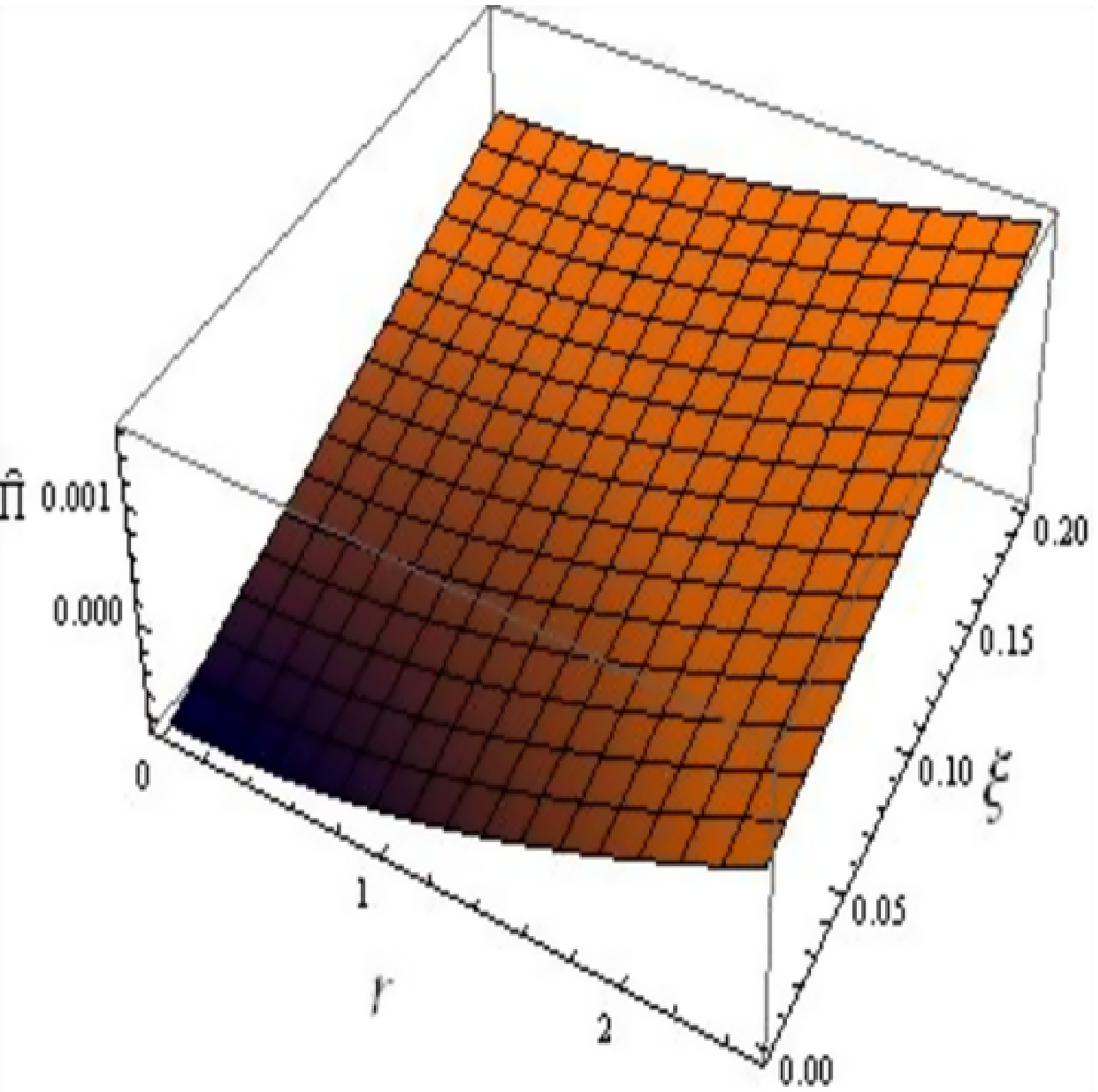,width=0.4\linewidth}
\caption{Plots of
$\hat{\mu},~\hat{\mathcal{P}}_{r},~\hat{\mathcal{P}}_{\bot}$ and
$\hat{\Pi}$ versus $r$ and $\xi$ with $\breve{M}=1M_{\bigodot}$ and
$H=(0.2)^{-1}M_{\bigodot}$ for Solution-I.}
\end{figure}
\begin{figure}\center
\epsfig{file=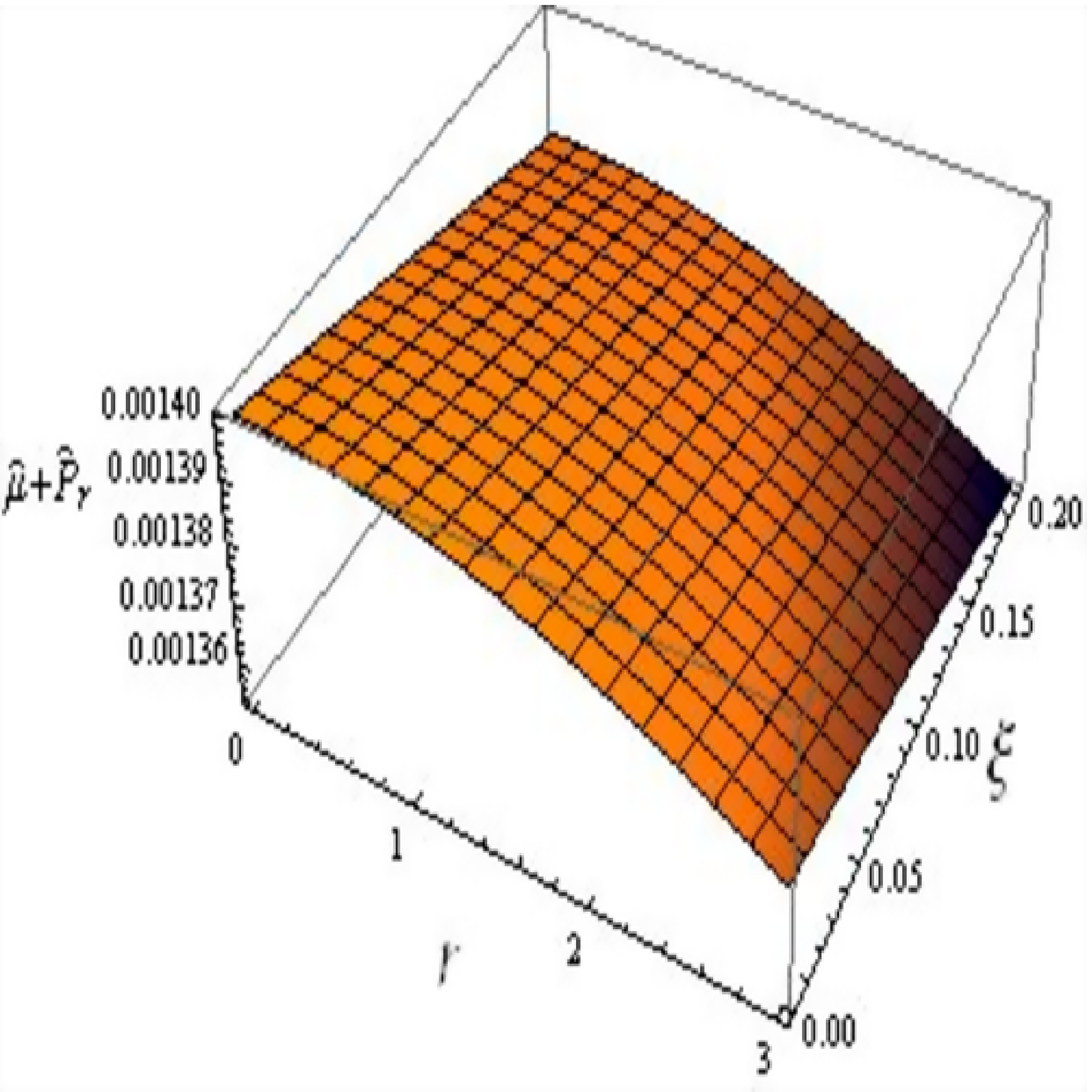,width=0.4\linewidth}\epsfig{file=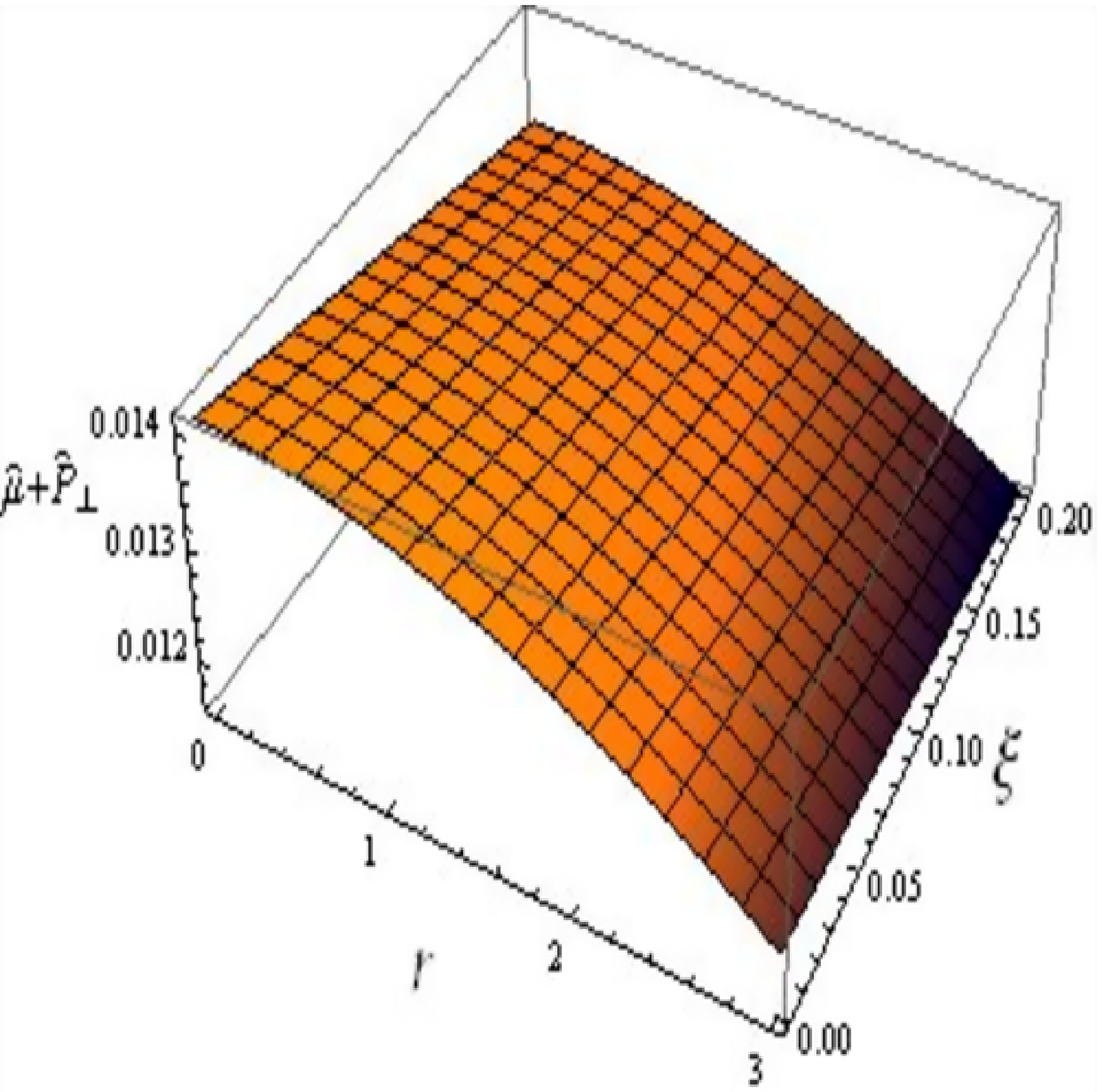,width=0.4\linewidth}
\epsfig{file=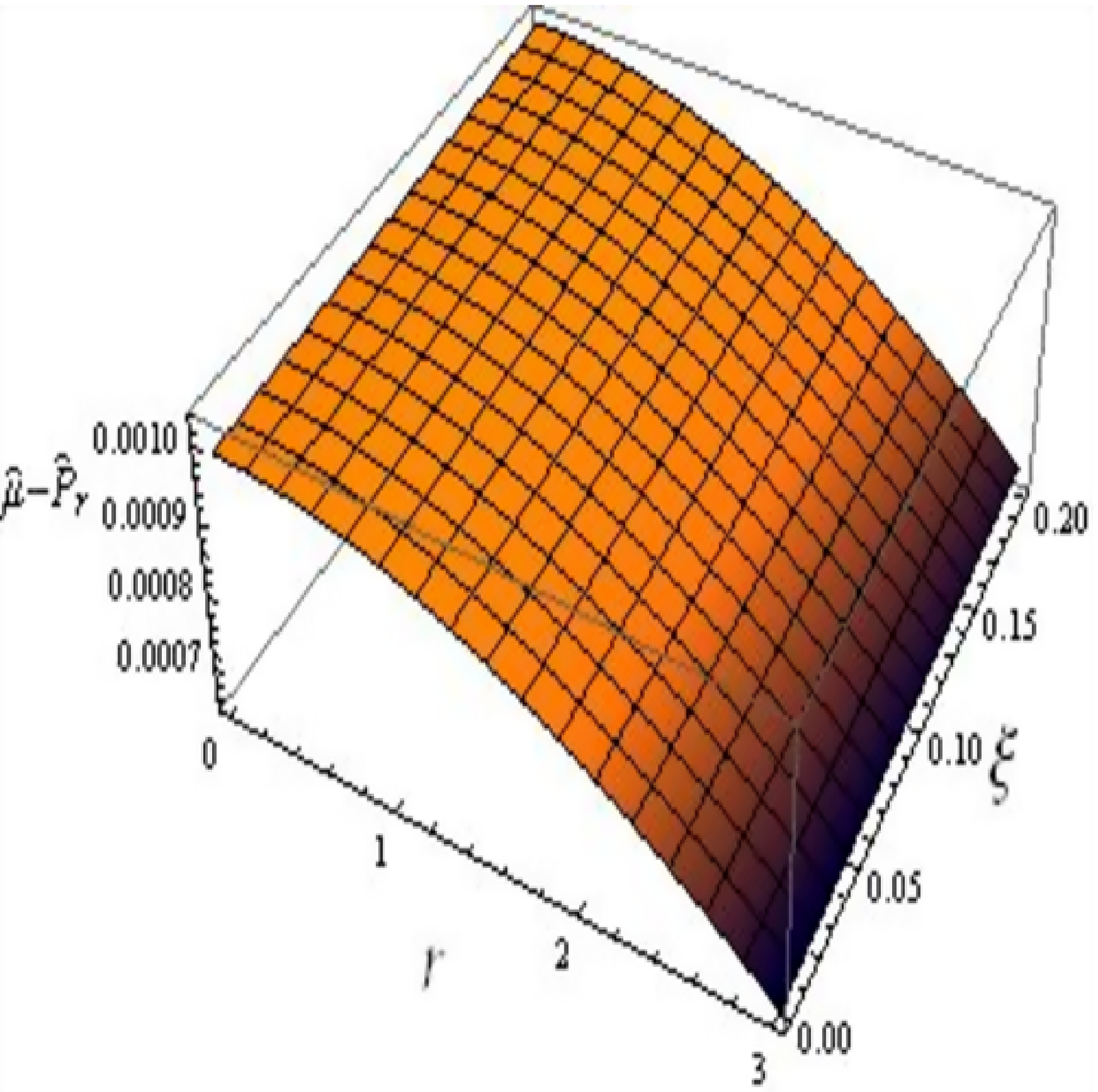,width=0.4\linewidth}\epsfig{file=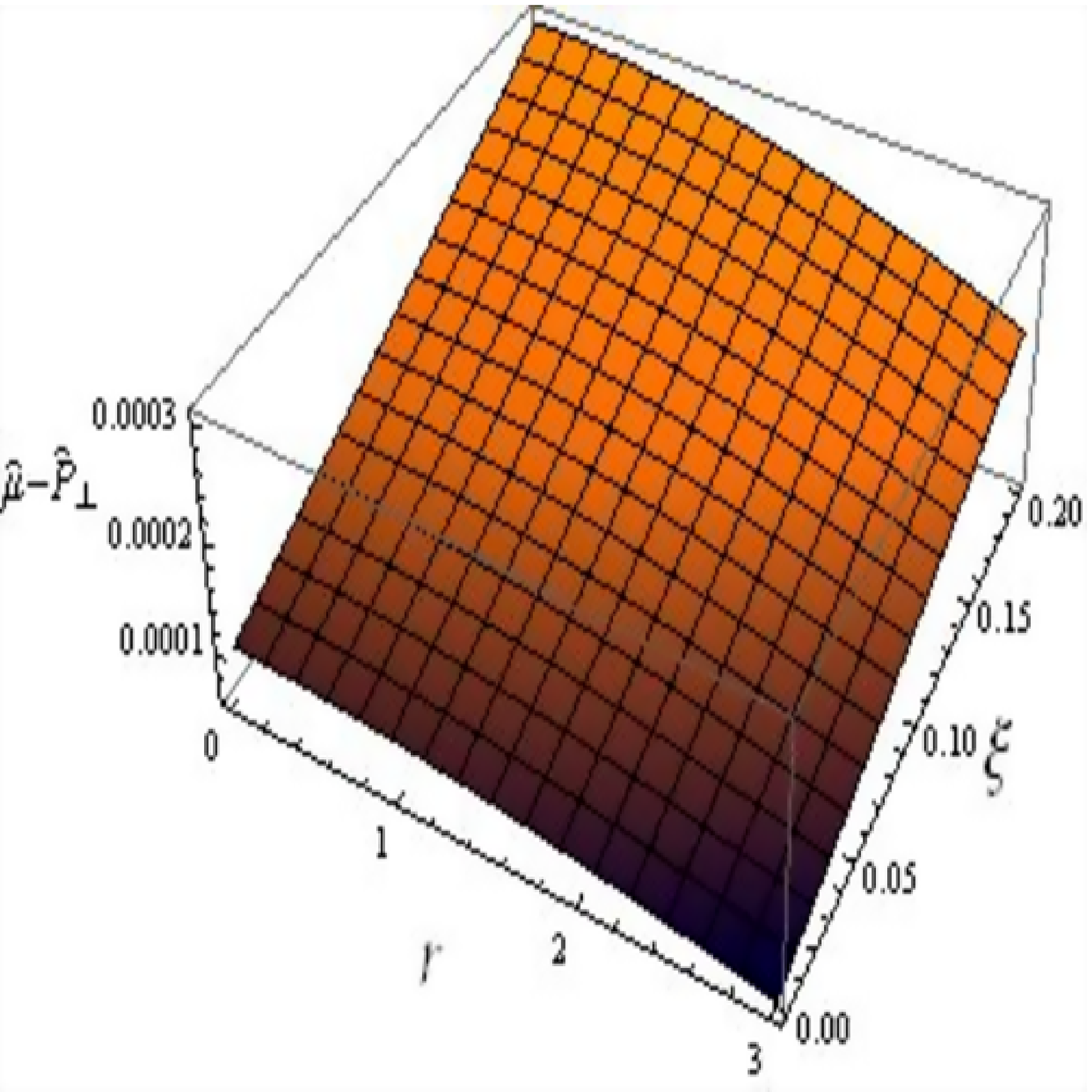,width=0.4\linewidth}
\epsfig{file=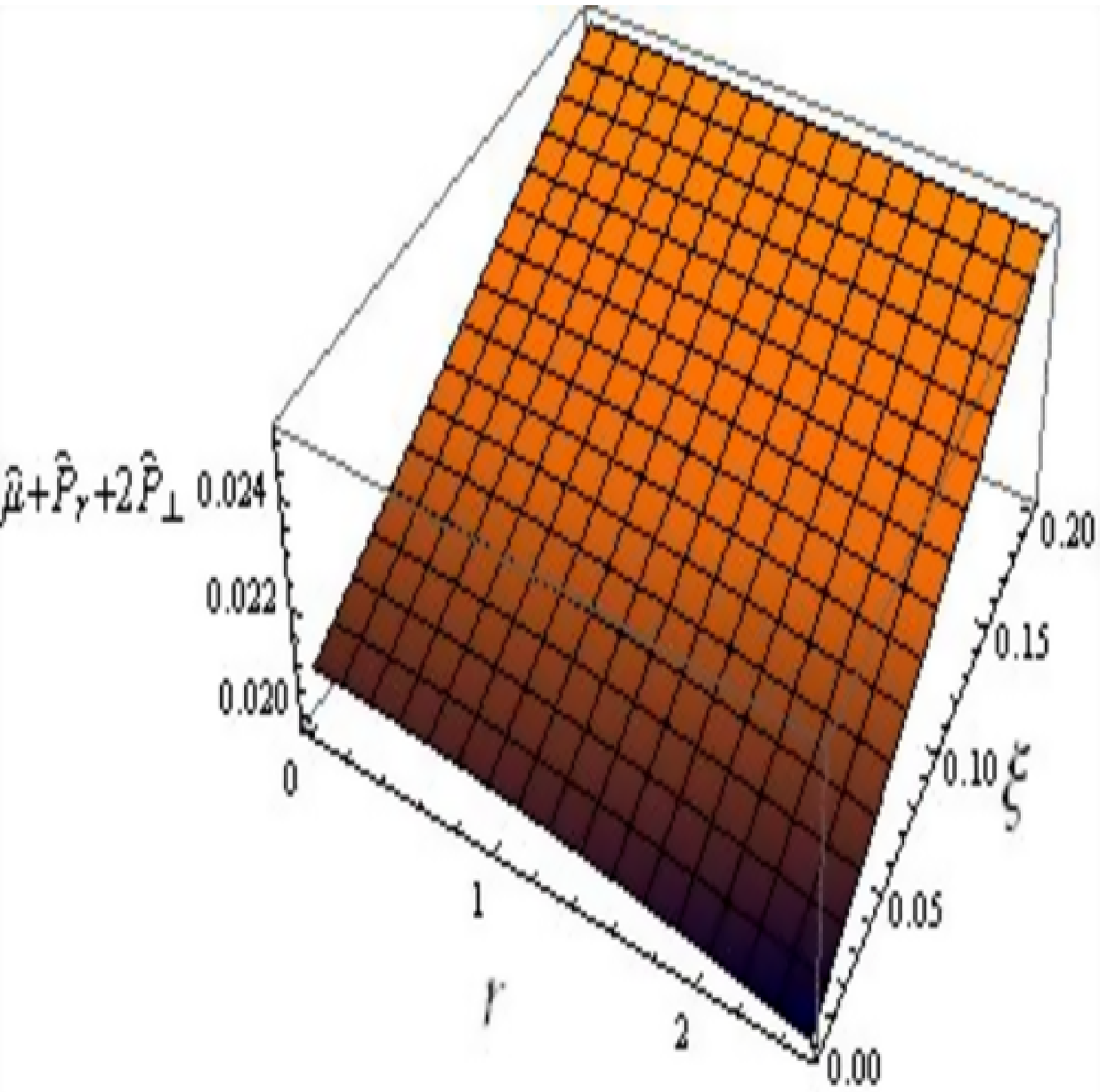,width=0.4\linewidth} \caption{Plots of
energy conditions versus $r$ and $\xi$ with
$\breve{M}=1M_{\bigodot}$ and $H=(0.2)^{-1}M_{\bigodot}$ for
Solution-I.}
\end{figure}

In the core of a celestial object, the physical variables (such as
pressure and energy density) should have finite, maximum and
positive values, while their behavior must decrease monotonically
towards the surface of star. Figure \textbf{2} (upper left) shows
that the effective energy density gains its maximum value in the
middle and decreases by increasing $r$. The effective energy density
enhances linearly with the increase of decoupling parameter, thus
larger is the decoupling parameter, more dense is the star. Figure
\textbf{2} also shows that the plots of radial and tangential
pressures are similar for the parameter $\varrho$ and decrease with
rise in $r$ as well as $\xi$. The anisotropic factor $\hat{\Pi}$
vanishes at the center and increases with the increase in $\xi$
which reveals that the system contains stronger anisotropy due to
the decoupling parameter. By means of graphical interpretation with
different values of the coupling constant $\varrho$, we examine that
small negative values of $\varrho$ provide the compatible physical
variables, while its positive values show their counter behavior.
Figure \textbf{3} shows that all energy conditions \eqref{g50} are
satisfied, hence our first solution is physically viable. It is
important to mention here that our Solution-I provides stable
configuration throughout the system, as shown by Figure \textbf{4}.
\begin{figure}\center
\epsfig{file=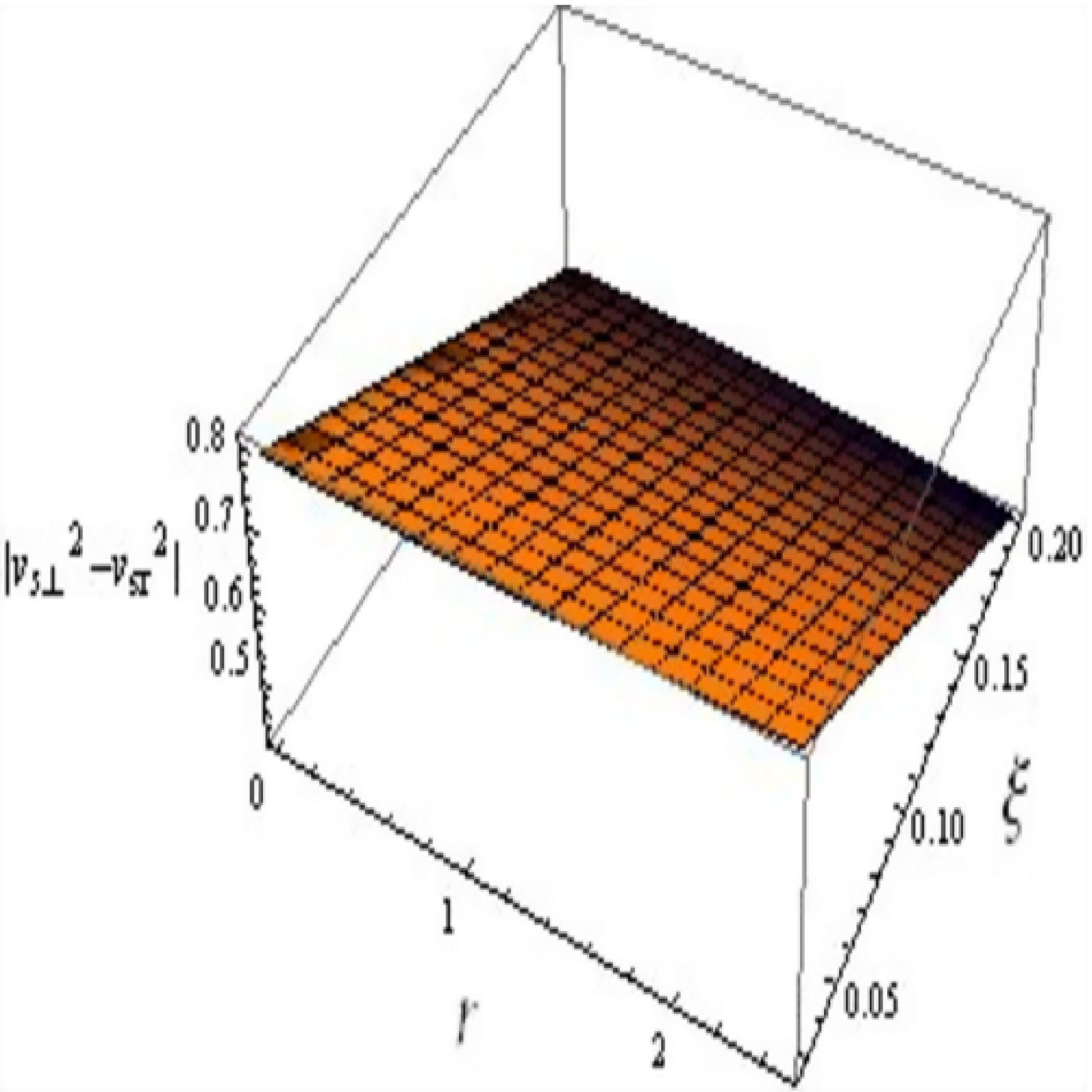,width=0.4\linewidth}\epsfig{file=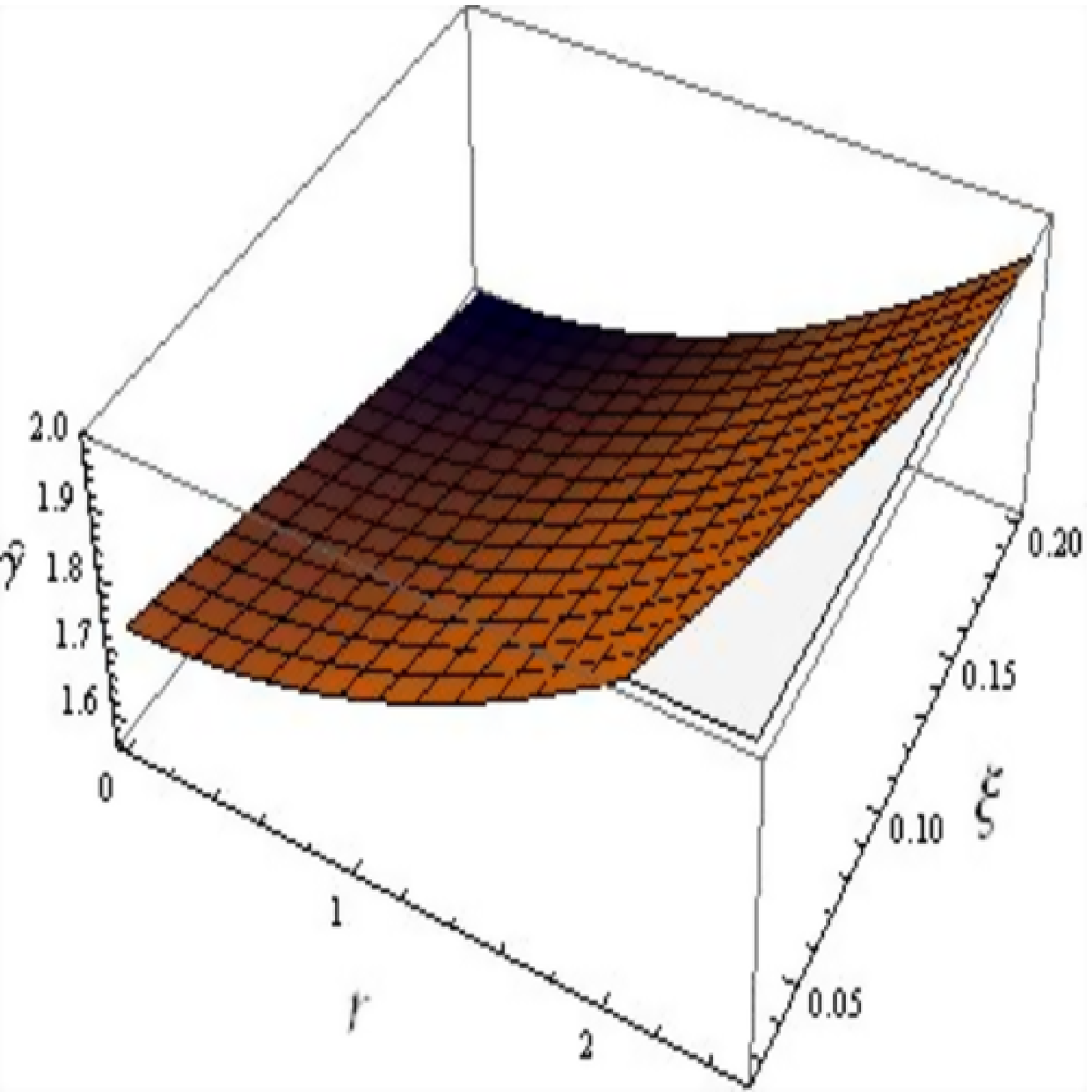,width=0.4\linewidth}
\caption{Plots of $|v_{s\perp}^2-v_{sr}^2|$ and adiabatic index
versus $r$ and $\xi$ with $\breve{M}=1M_{\bigodot}$ and
$H=(0.2)^{-1}M_{\bigodot}$ for Solution-I.}
\end{figure}
\begin{figure}\center
\epsfig{file=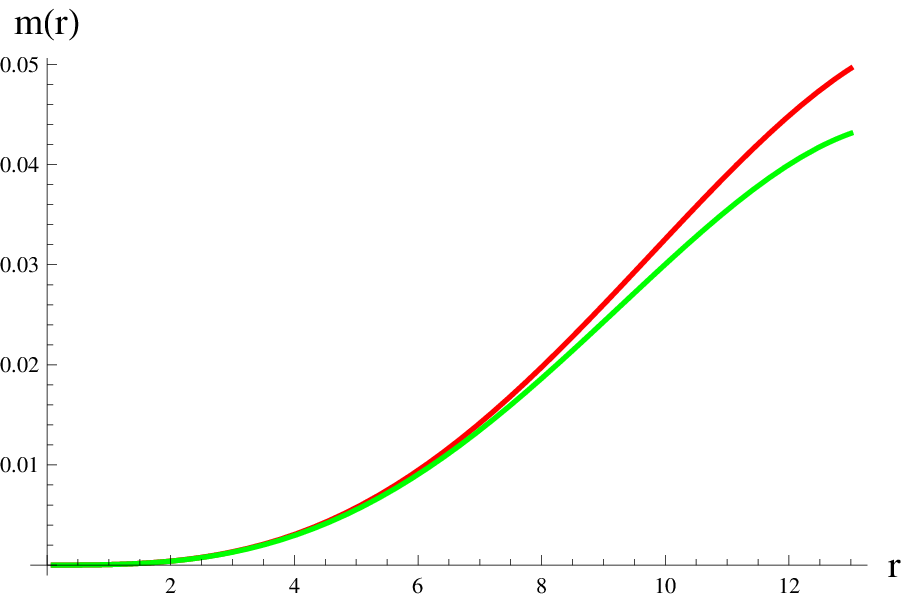,width=0.4\linewidth}\epsfig{file=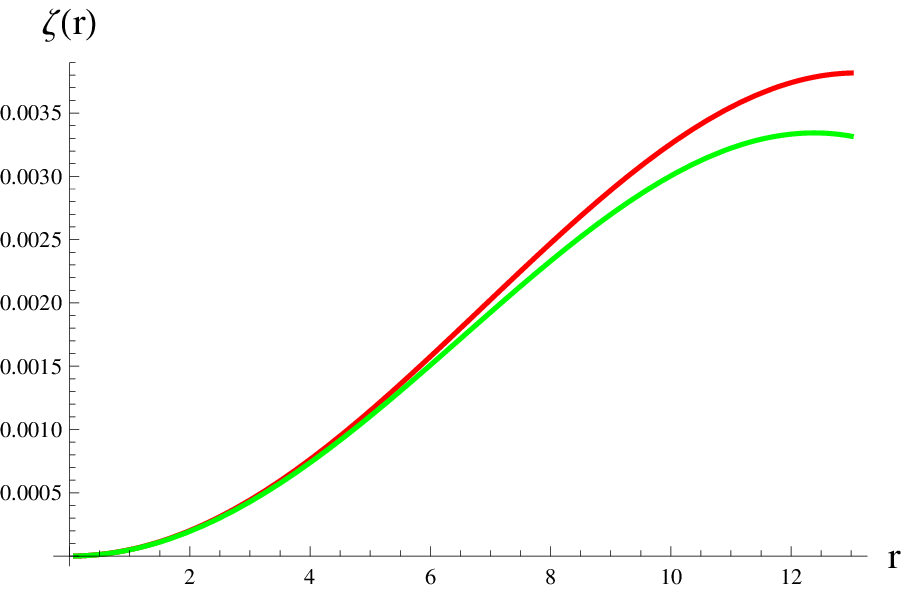,width=0.4\linewidth}
\epsfig{file=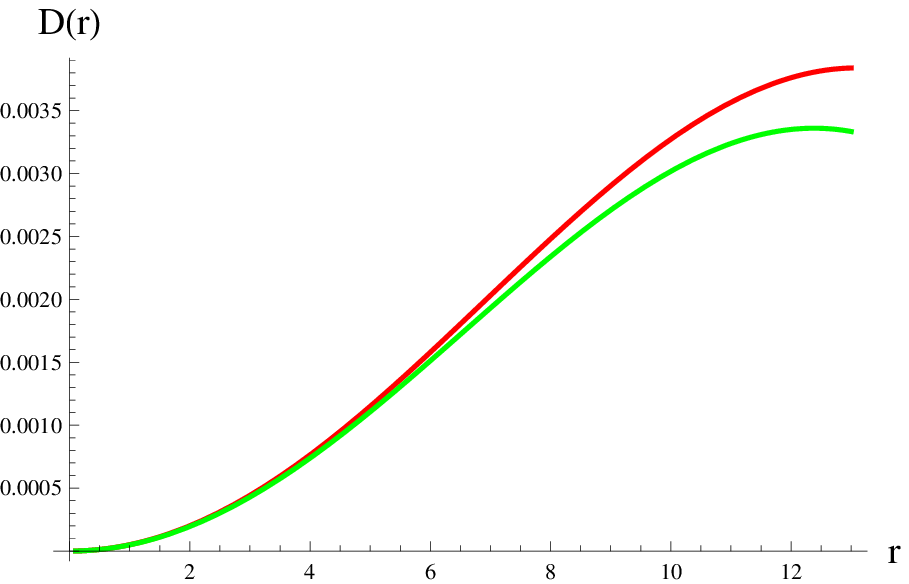,width=0.4\linewidth} \caption{Plots of
mass, compactness and redshift parameters corresponding to
$\xi=0.10$ (red) and $\xi=0.14$ (green) for Solution-II.}
\end{figure}
\begin{figure}\center
\epsfig{file=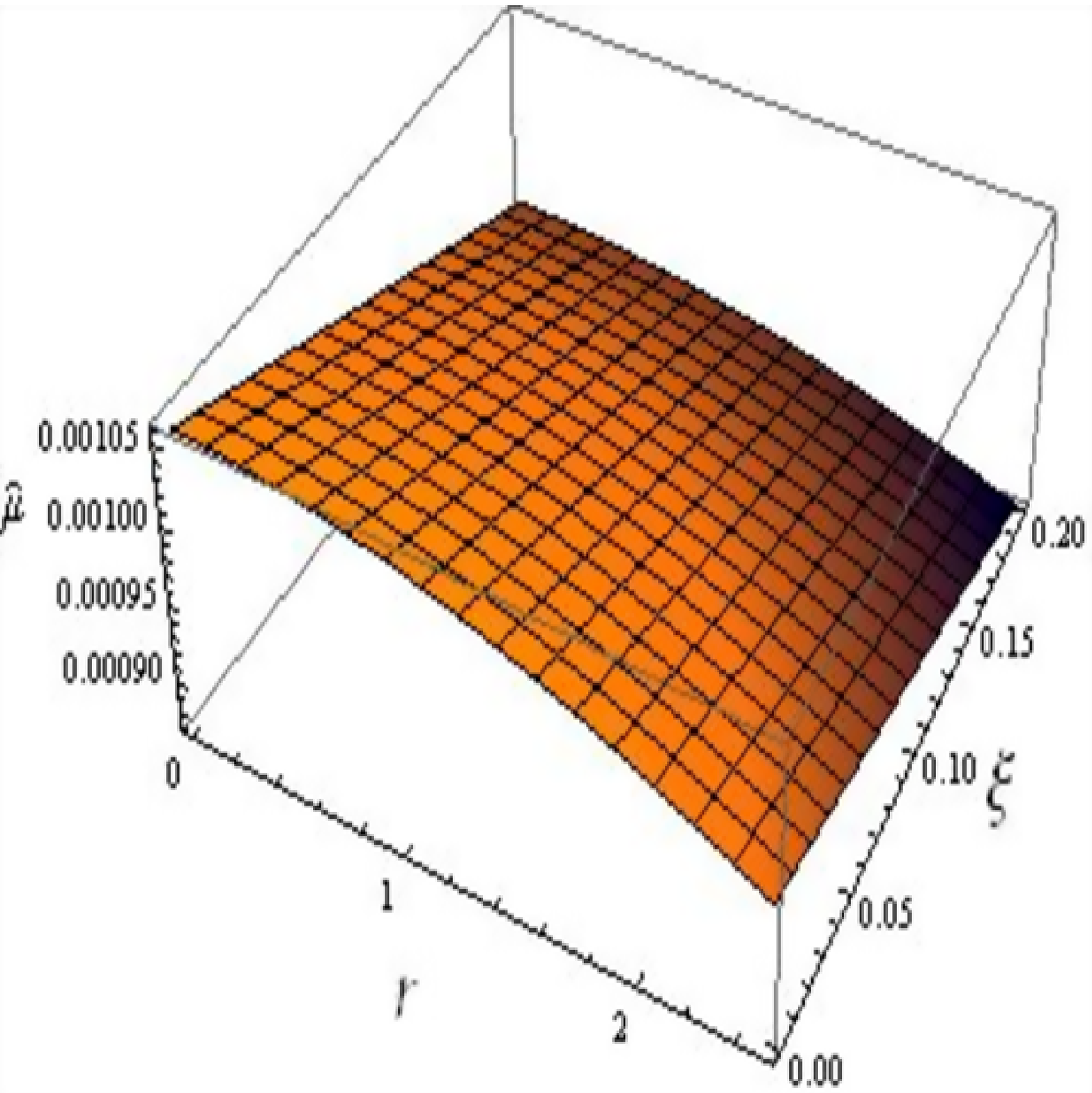,width=0.4\linewidth}\epsfig{file=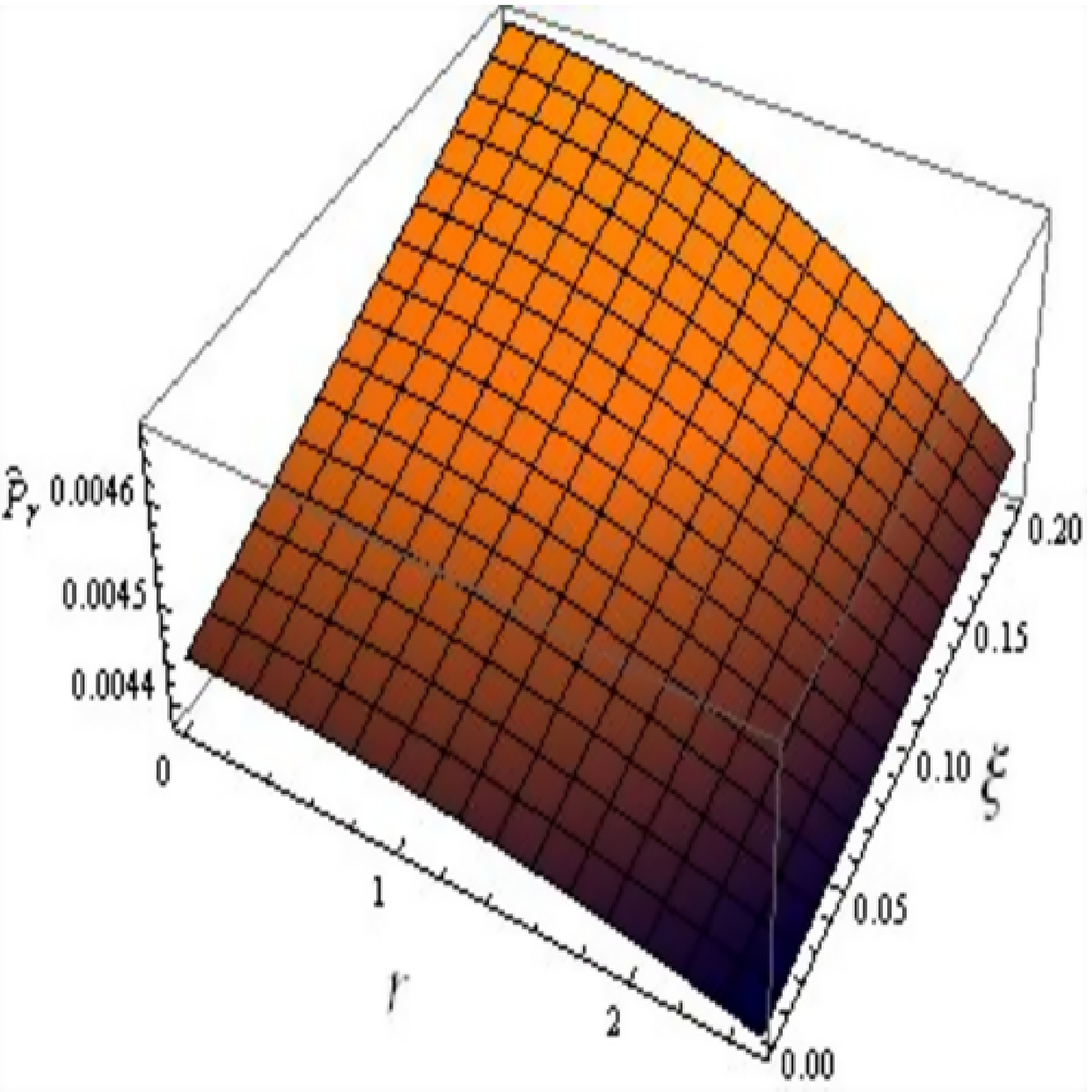,width=0.4\linewidth}
\epsfig{file=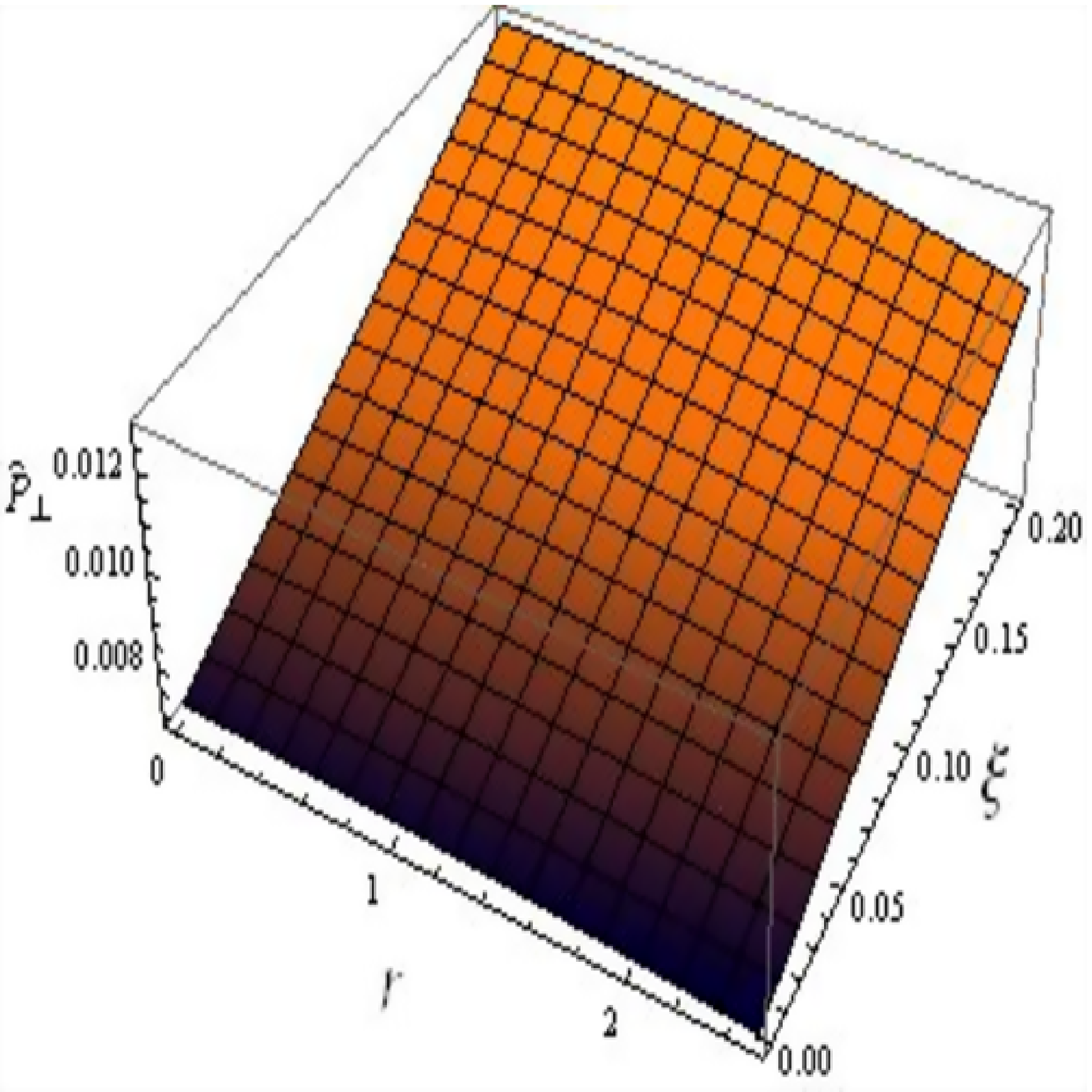,width=0.4\linewidth}\epsfig{file=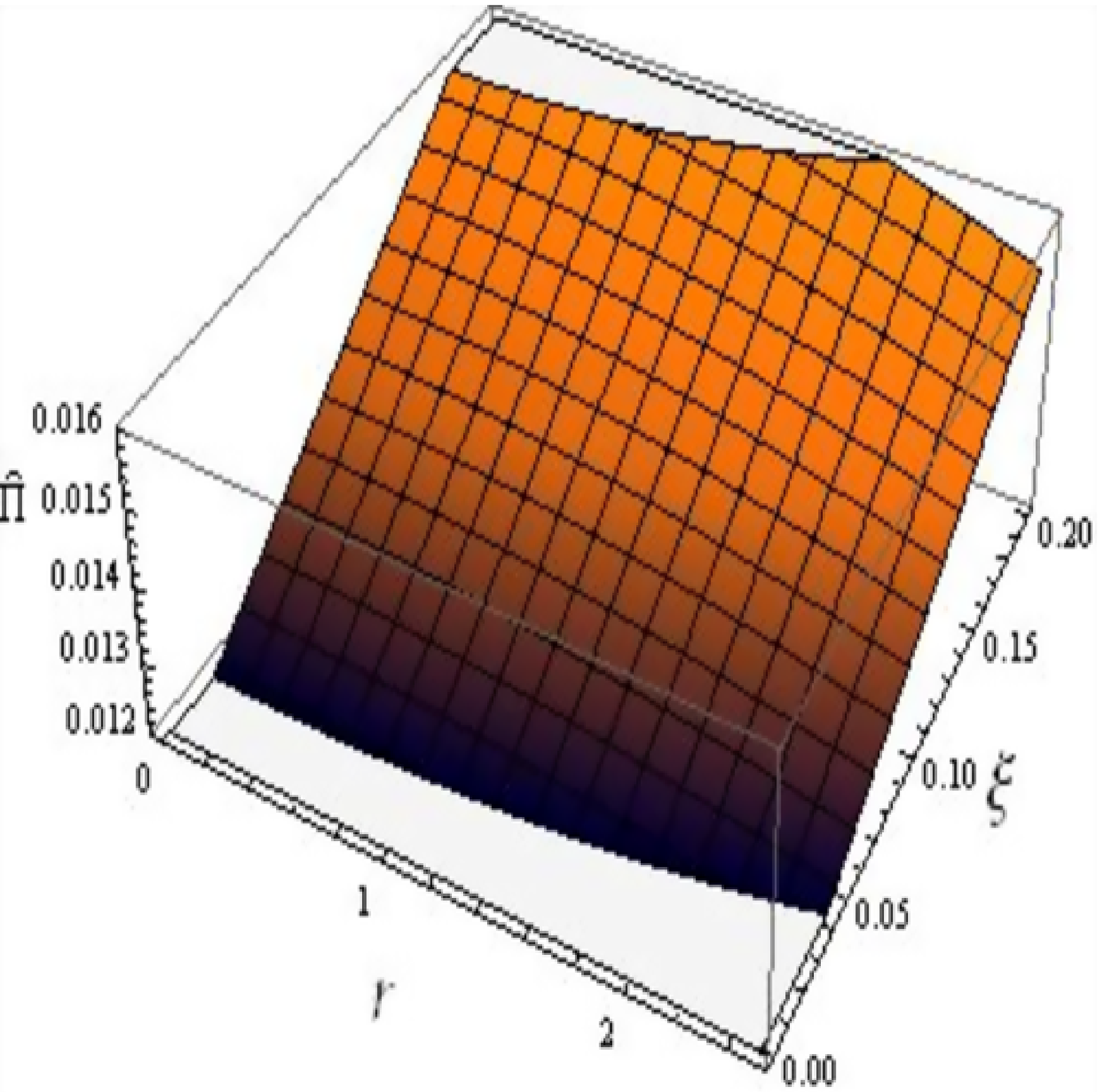,width=0.4\linewidth}
\caption{Plots of
$\hat{\mu},~\hat{\mathcal{P}}_{r},~\hat{\mathcal{P}}_{\bot}$ and
$\hat{\Pi}$ versus $r$ and $\xi$ with $\breve{M}=1M_{\bigodot}$ and
$H=(0.2)^{-1}M_{\bigodot}$ for Solution-II.}
\end{figure}
\begin{figure}\center
\epsfig{file=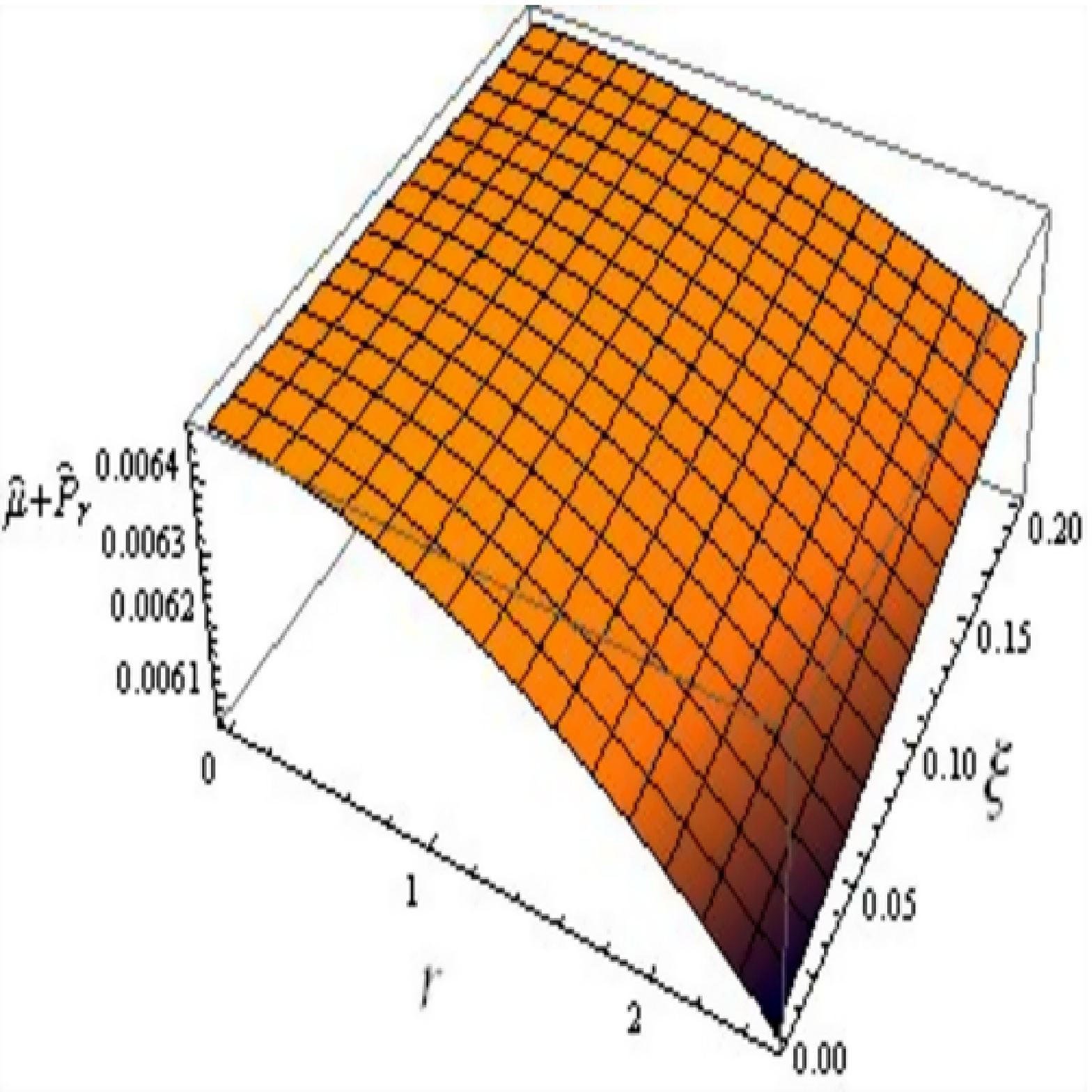,width=0.4\linewidth}\epsfig{file=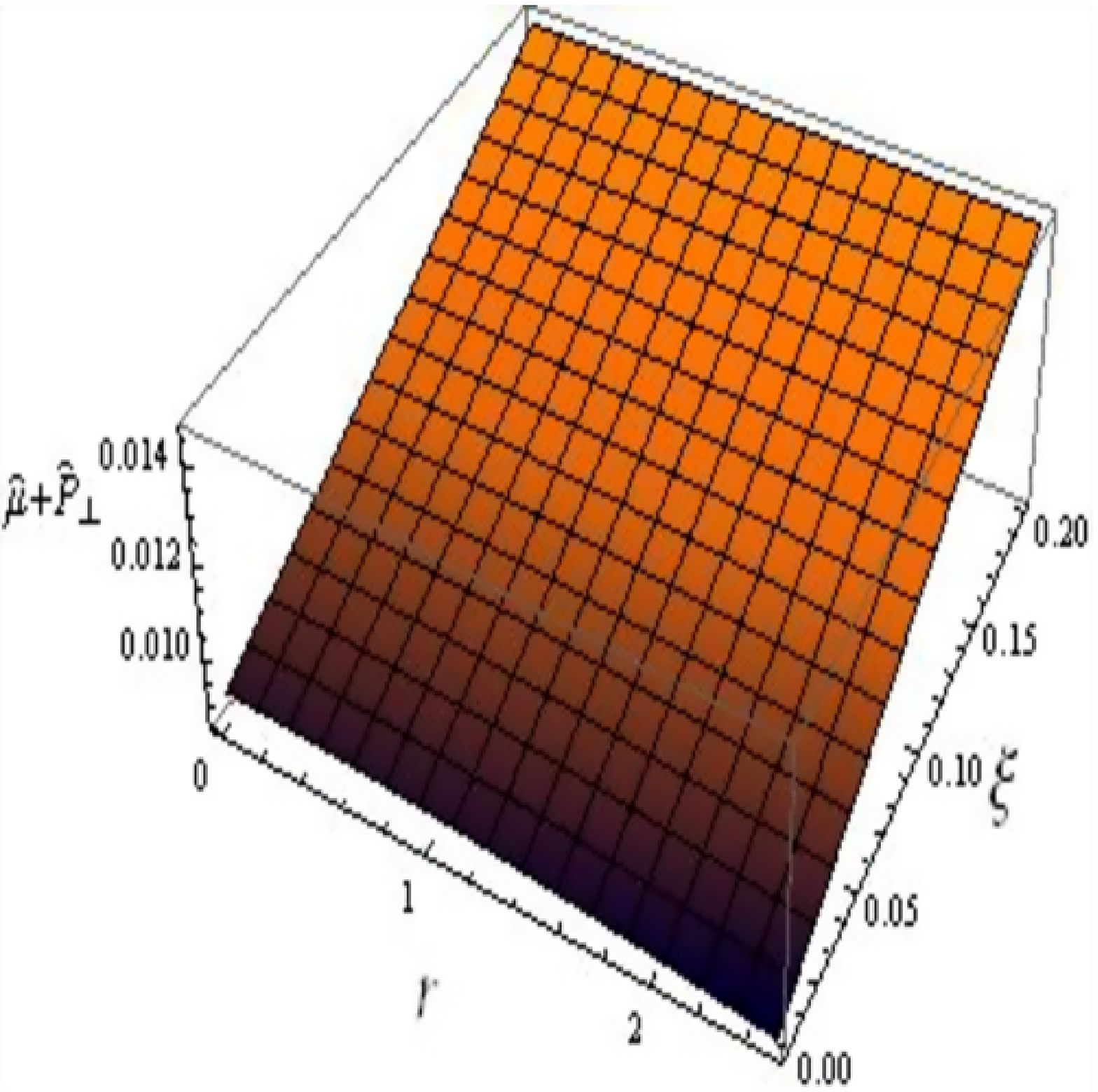,width=0.4\linewidth}
\epsfig{file=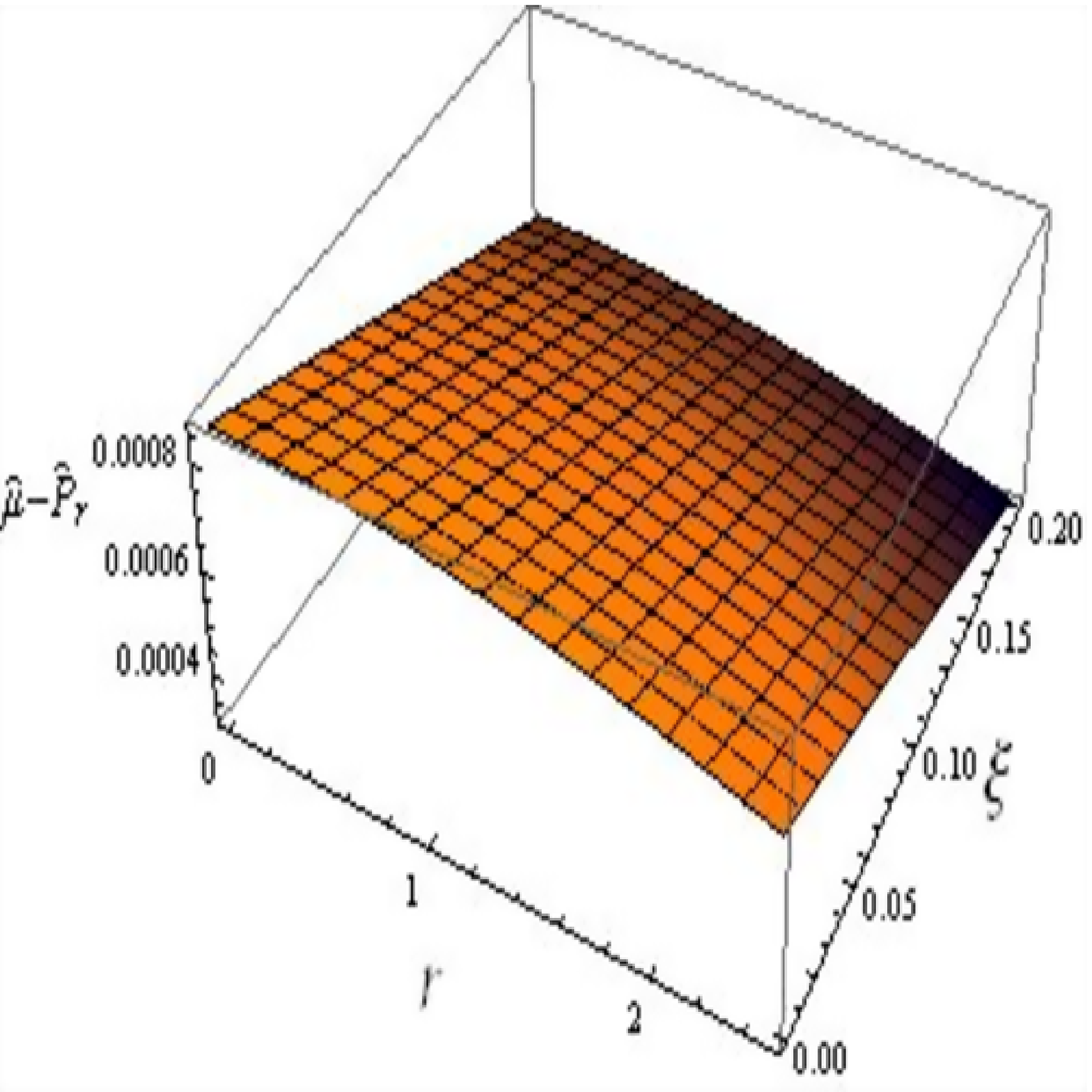,width=0.4\linewidth}\epsfig{file=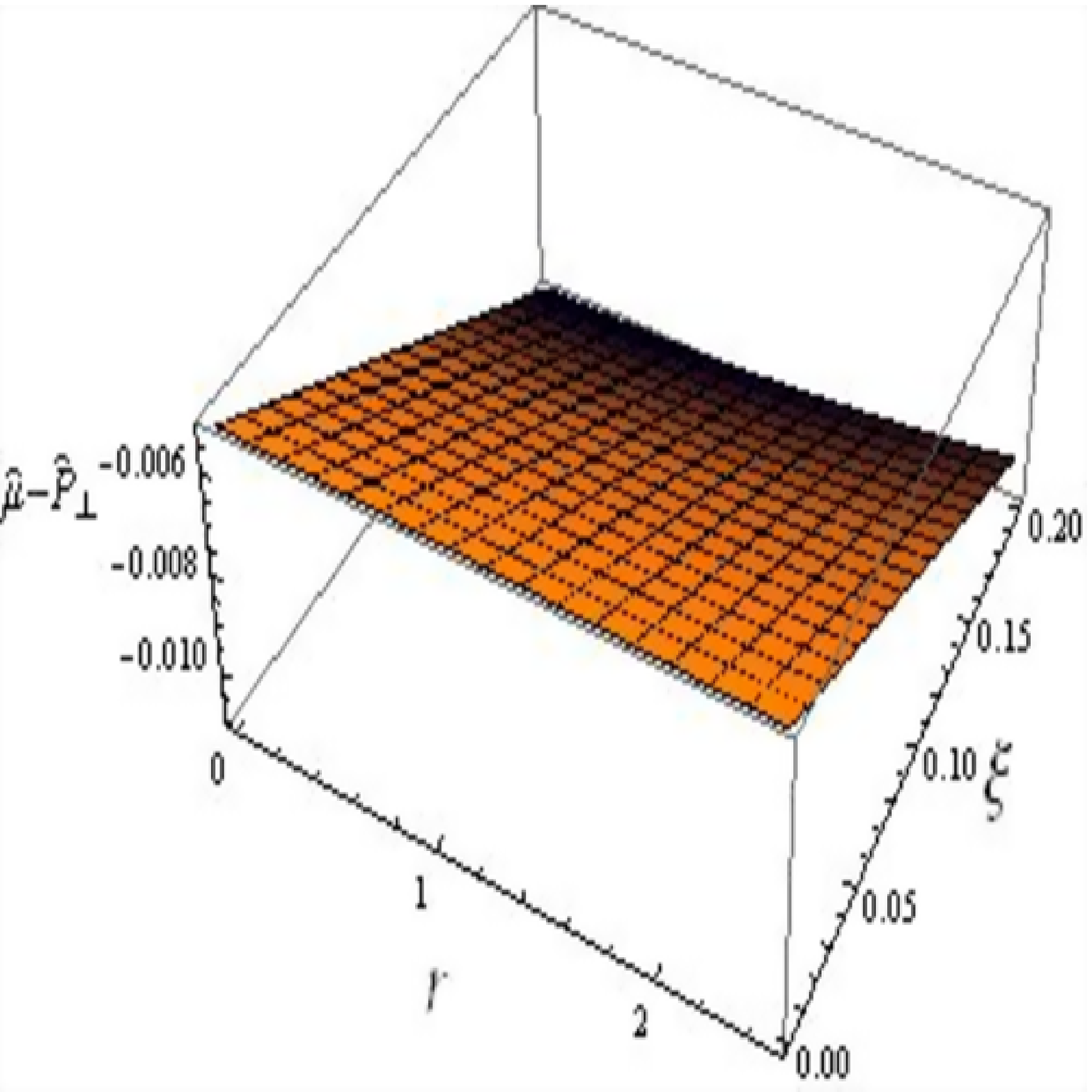,width=0.4\linewidth}
\epsfig{file=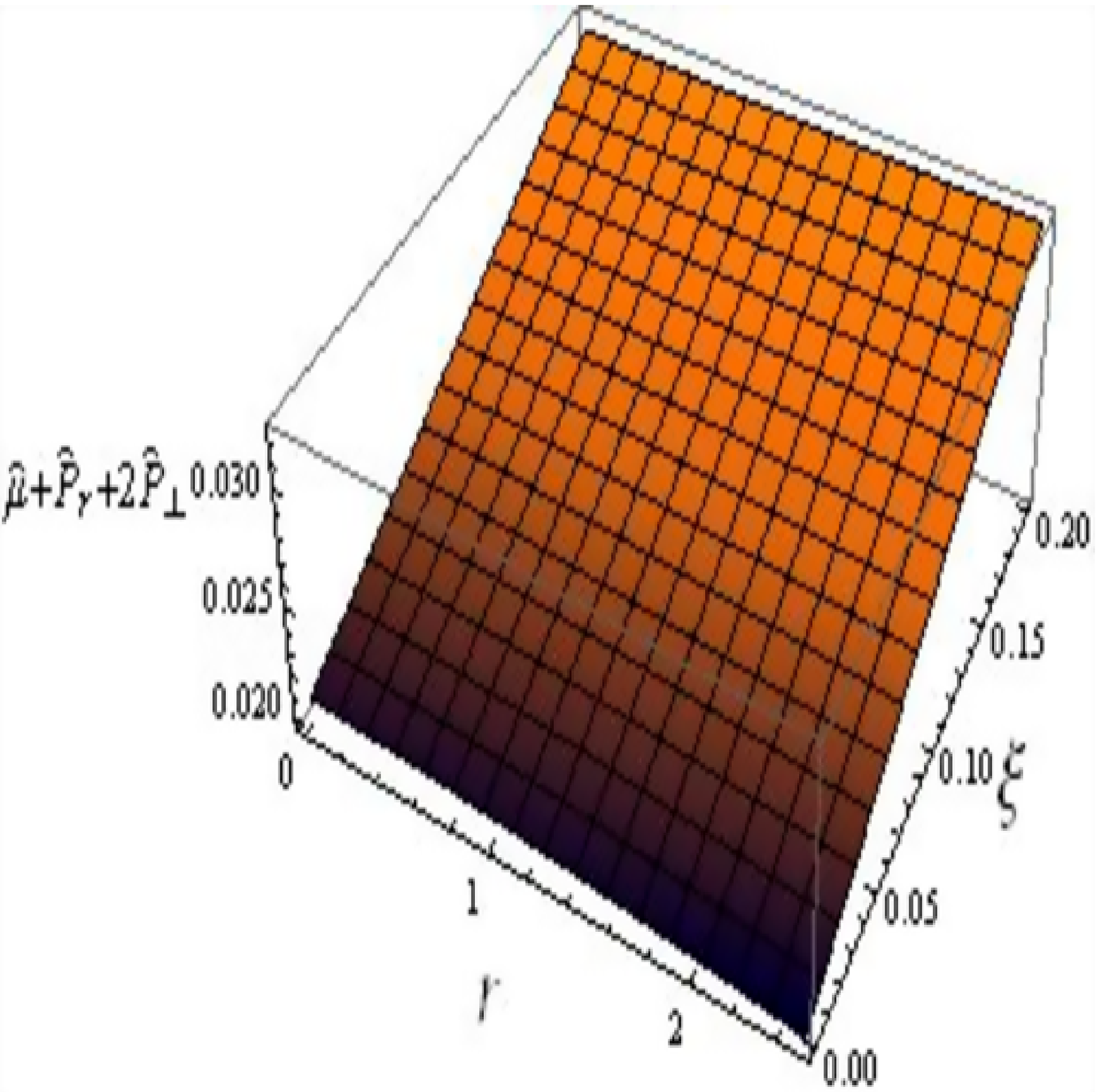,width=0.4\linewidth} \caption{Plots of
energy conditions versus $r$ and $\xi$ with
$\breve{M}=1M_{\bigodot}$ and $H=(0.2)^{-1}M_{\bigodot}$ for
Solution-II.}
\end{figure}
\begin{figure}\center
\epsfig{file=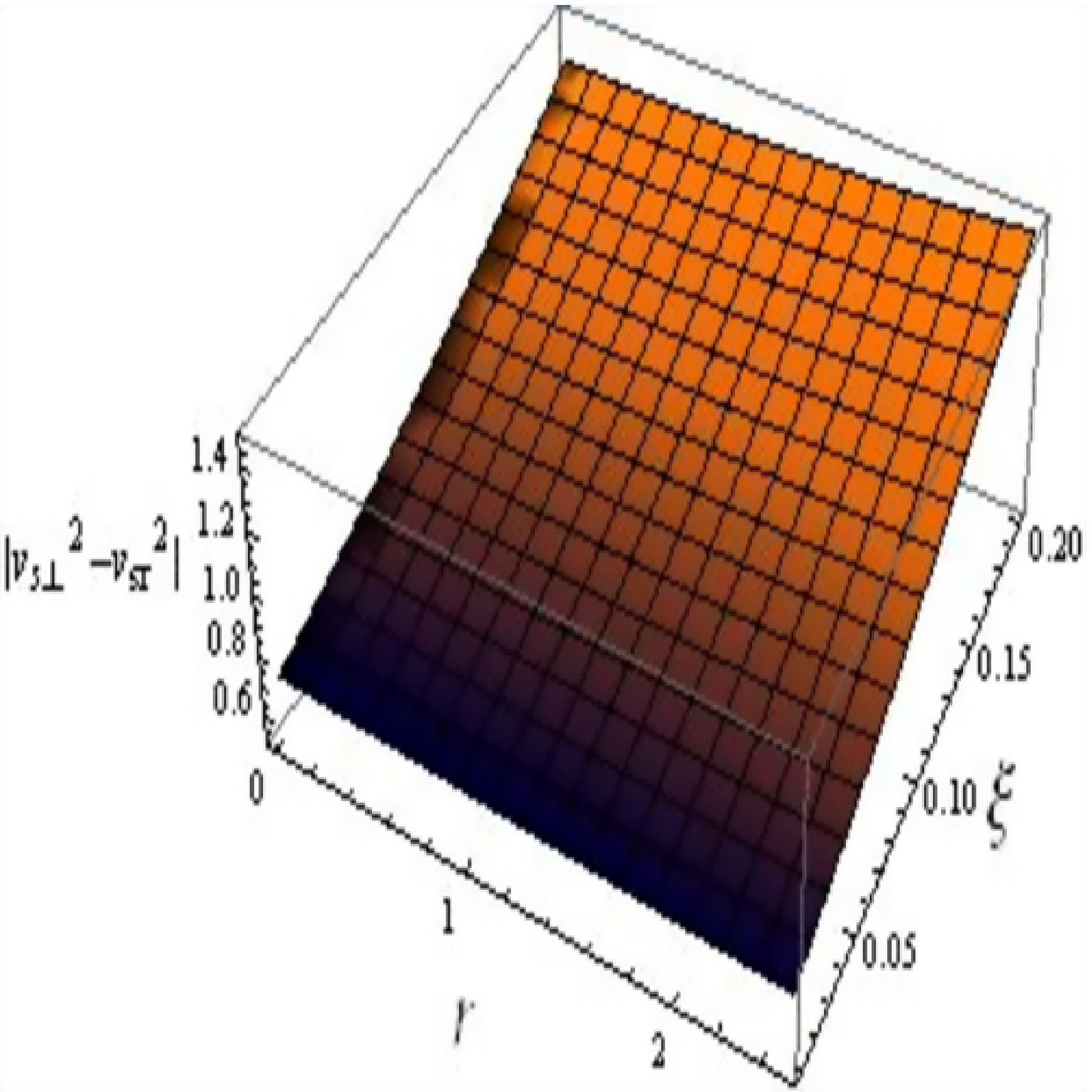,width=0.4\linewidth}\epsfig{file=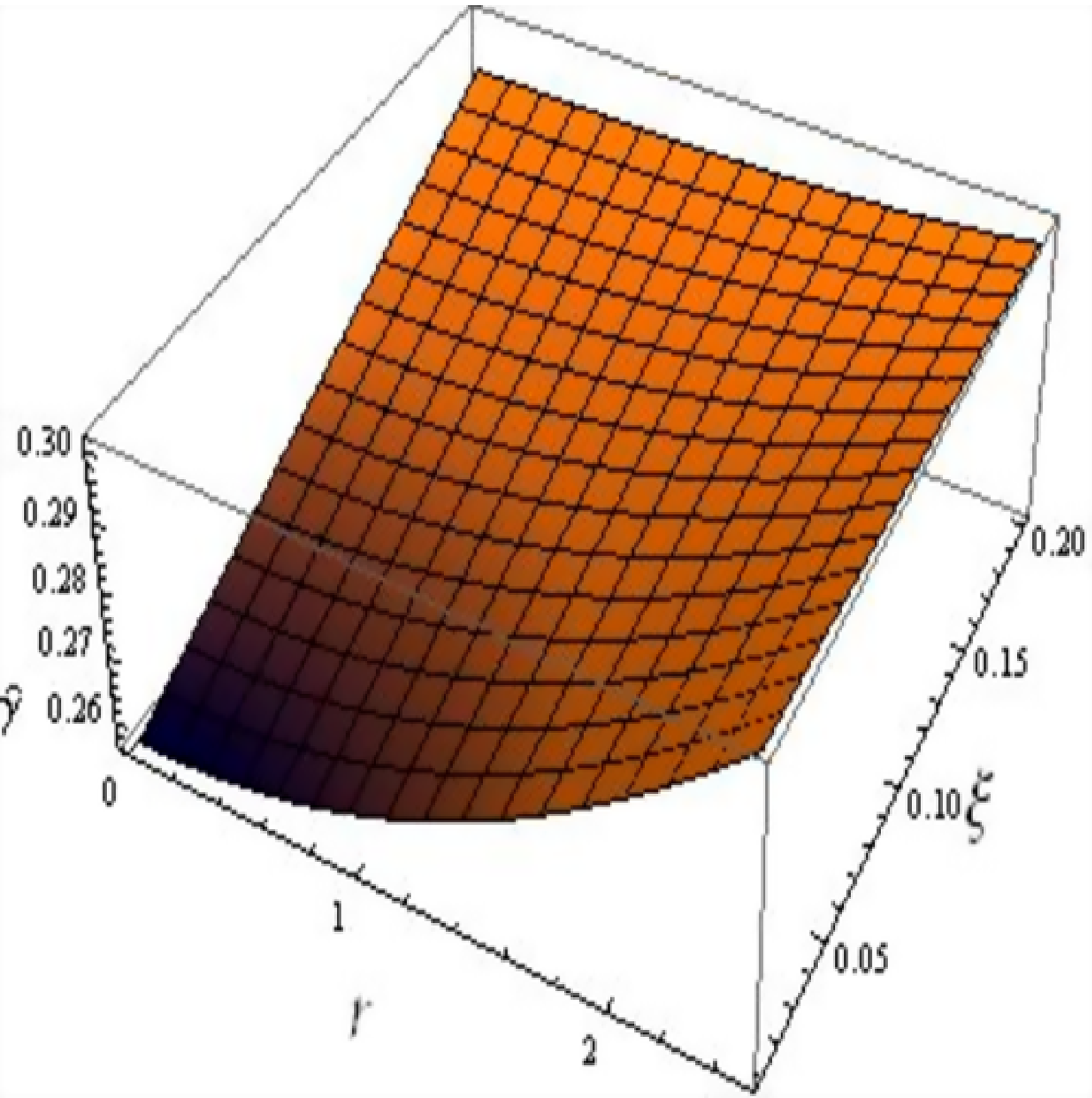,width=0.4\linewidth}
\caption{Plots of $|v_{s\perp}^2-v_{sr}^2|$ and adiabatic index
versus $r$ and $\xi$ with $\breve{M}=1M_{\bigodot}$ and
$H=(0.2)^{-1}M_{\bigodot}$ for Solution-II.}
\end{figure}

The physical features of our developed Solution-II are now being
examined by taking $\varrho=-0.05$. The constants $X$ and $Z$ are
shown in equations \eqref{g37} and \eqref{g56}. Figure \textbf{5}
(upper left) shows decreasing behavior of the mass of sphere
\eqref{g6} with increasing $\xi$. It can also be seen from the same
figure (upper right and lower) that the compactness $(\zeta(r))$ and
redshift $(D(r))$ comply with their required limits. The behavior of
material variables and anisotropy is illustrated in Figure
\textbf{6}. The effective energy density decreases with rise in the
decoupling parameter whereas both the effective pressures increase
linearly. Figure \textbf{6} (lower right) discloses that the
behavior of anisotropy is monotonically increasing with respect to
$\xi$. Our second solution fulfills all energy conditions
\eqref{g50} except $\hat{\mu}-\hat{\mathcal{P}}_{\bot} \geq 0$, as
shown in Figure \textbf{7}. Hence, it is not physically viable.
Figure \textbf{8} (right) confirms that the Solution-II is unstable
for all values of parameter $\xi$ as it does not fulfill the
required limit of adiabatic index, i.e., it takes value less than
$\frac{4}{3}$ throughout.

\section{Conclusions}

This paper is intended to examine two different anisotropic
solutions for spherically symmetric spacetime for a linear model
$\mathcal{R}+\varrho \mathcal{Q}$ in
$f(\mathcal{R},\mathcal{T},\mathcal{Q})$ gravity. We have utilized
EGD technique to find such solutions. We have introduced an
additional source $\Delta_{\gamma\chi}$ in the modified action and
obtained anisotropic field equations which have further been divided
into two sets by employing deformation functions. We have assumed
the isotropic Krori-Barua ansatz to deal with the first set in this
gravity involving three unknowns which have been calculated through
junction conditions. The second sector \eqref{g21}-\eqref{g23}
involves five unknowns, thus we have employed two constraints to
close the system, one of them is the equation of state
$\Delta^{0}_{0}=\alpha\Delta^{1}_{1}+\beta\Delta^{2}_{2}$ which
depends upon anisotropic components. The other constraint has been
taken as pressure-like or density-like which has led to the
Solutions-I and II, respectively.

In order to check the feasibility of both solutions corresponding to
the decoupling parameter, we have checked the graphs of effective
forms of physical variables
$(\hat{\mu},\hat{\mathcal{P}}_{r},\hat{\mathcal{P}}_{\bot})$,
pressure anisotropy $(\hat{\Pi})$ as well as energy bounds
\eqref{g50} by taking $\varrho=-0.1$ and $-0.05$. The redshift and
compactness for both solutions have been found to be in their
respective limits. The self-gravitating geometry \eqref{g6} becomes
more massive for Solution-I as the decoupling parameter $\xi$
increases, while the mass decreases in case of the Solution-II. We
have also explored the stability of these solutions through two
approaches. The first solution provides viable as well as stable
geometry, while the sphere \eqref{g6} does not meet the criteria of
viability and stability for Solution-II. We summarize that the
behavior of these modified solutions are in contrast with GR
\cite{40,41} and $f(\mathcal{G})$ theory \cite{41a}, as Solution-I
shows more stable behavior while Solution-II is not stable
throughout. The resulting Solution-I is also not compatible with
\cite{35} as that solution is not stable throughout the considered
range of the parameter $\xi$, i.e., $(0,0.28)$. Finally, these
results can be reduced to GR by considering
$f(\mathcal{R},\mathcal{T},\mathcal{Q})=\mathcal{R}$ in the action
\eqref{g1}.

\vspace{0.25cm}

\section*{Appendix A}

The modified corrections appear in equations \eqref{g8}-\eqref{g10}
are given as
\begin{eqnarray}\nonumber
\mathcal{T}_{0}^{0(\mathcal{D})}&=&\frac{1}{f_{\mathcal{R}}+\mu
f_{\mathcal{Q}}}\left[\mu\left\{f_{\mathcal{Q}}\left(\frac{\psi'^2}{2e^{\phi}}-\frac{\psi'}{re^{\phi}}+\frac{\psi'\phi'}{4e^{\phi}}
-\frac{\psi''}{2e^{\phi}}-\frac{1}{2}\mathcal{R}\right)+f'_{\mathcal{Q}}\left(\frac{\psi'}{2e^{\phi}}\right.\right.\right.\\\nonumber
&-&\left.\left.\frac{\phi'}{4e^{\phi}}+\frac{1}{re^{\phi}}\right)+\frac{f''_{\mathcal{Q}}}{2e^{\phi}}-2f_{\mathcal{T}}\right\}
+\mu'\left\{f_{\mathcal{Q}}\left(\frac{\psi'}{2e^{\phi}}
+\frac{1}{re^{\phi}}-\frac{\phi'}{4e^{\phi}}\right)+\frac{f'_{\mathcal{Q}}}{e^{\phi}}\right\}\\\nonumber
&+&\frac{f_{\mathcal{Q}}\mu''}{2e^{\phi}}+\mathcal{P}\left\{f_{\mathcal{Q}}
\left(\frac{3\phi'^2}{4e^{\phi}}-\frac{2}{r^2e^{\phi}}-\frac{\phi''}{2e^{\phi}}\right)
-f'_{\mathcal{Q}}\left(\frac{5\phi'}{4e^{\phi}}-\frac{1}{re^{\phi}}\right)+\frac{f''_{\mathcal{Q}}}{2e^{\phi}}\right\}\\\nonumber
&+&\mathcal{P}'\left\{f_{\mathcal{Q}}\left(\frac{1}{re^{\phi}}
-\frac{5\phi'}{4e^{\phi}}\right)+\frac{f'_{\mathcal{Q}}}{e^{\phi}}\right\}
+\frac{f_{\mathcal{Q}}\mathcal{P}''}{2e^{\phi}}+\frac{\mathcal{R}f_{\mathcal{R}}}{2}
+f'_{\mathcal{R}}\left(\frac{\phi'}{2e^{\phi}}-\frac{2}{re^{\phi}}\right)\\\nonumber
&-&\left.\frac{f''_{\mathcal{R}}}{e^{\phi}}-\frac{f}{2}\right],\\\nonumber
\mathcal{T}_{1}^{1(\mathcal{D})}&=&\frac{1}{f_{\mathcal{R}}+\mu
f_{\mathcal{Q}}}\left[\mu\left(f_{\mathcal{T}}-\frac{f_{\mathcal{Q}}\psi'^2}{4e^{\phi}}
+\frac{f'_{\mathcal{Q}}\psi'}{4e^{\phi}}\right)+\frac{f_{\mathcal{Q}}\mu'\psi'}{4e^{\phi}}+\mathcal{P}\left\{f_{\mathcal{T}}
+f_{\mathcal{Q}}\left(\frac{\psi''}{e^{\phi}}\right.\right.\right.\\\nonumber
&-&\left.\left.\frac{\phi'^2}{e^{\phi}}+\frac{\psi'^2}{2e^{\phi}}-\frac{3\psi'\phi'}{4e^{\phi}}-\frac{3\phi'}{re^{\phi}}
+\frac{2}{r^2e^{\phi}}+\frac{1}{2}\mathcal{R}\right)
-f'_{\mathcal{Q}}\left(\frac{\psi'}{4e^{\phi}}+\frac{2}{re^{\phi}}\right)\right\}\\\nonumber
&-&\left.\mathcal{P}'f_{\mathcal{Q}}\left(\frac{\psi'}{4e^{\phi}}+\frac{2}{re^{\phi}}\right)+\frac{f}{2}-\frac{\mathcal{R}f_{\mathcal{R}}}{2}
-f'_{\mathcal{R}}\left(\frac{\psi'}{2e^{\phi}}+\frac{2}{re^{\phi}}\right)\right],\\\nonumber
\mathcal{T}_{2}^{2(\mathcal{D})}&=&\frac{1}{f_{\mathcal{R}}+\mu
f_{\mathcal{Q}}}\left[\mu\left(f_{\mathcal{T}}-\frac{f_{\mathcal{Q}}\psi'^2}{4e^{\phi}}
+\frac{f'_{\mathcal{Q}}\psi'}{4e^{\phi}}\right)+\frac{f_{\mathcal{Q}}\mu'\psi'}{4e^{\phi}}+\mathcal{P}\left\{f_{\mathcal{T}}
+f_{\mathcal{Q}}\left(\frac{\phi''}{2e^{\phi}}\right.\right.\right.\\\nonumber
&+&\left.\frac{\psi'}{2re^{\phi}}-\frac{3\phi'^2}{4e^{\phi}}-\frac{\phi'}{2re^{\phi}}+\frac{1}{r^2e^{\phi}}-\frac{2}{r^2}
+\frac{1}{2}\mathcal{R}\right)+f'_{\mathcal{Q}}\left(\frac{3\phi'}{2e^{\phi}}-\frac{\psi'}{4e^{\phi}}-\frac{3}{re^{\phi}}\right)\\\nonumber
&-&\left.\frac{f''_{\mathcal{Q}}}{e^{\phi}}\right\}+\mathcal{P}'\left\{f_{\mathcal{Q}}\left(\frac{3\phi'}{2e^{\phi}}-\frac{\psi'}{4e^{\phi}}
-\frac{3}{re^{\phi}}\right)-\frac{2f'_{\mathcal{Q}}}{e^{\phi}}\right\}-\frac{f_{\mathcal{Q}}\mathcal{P}''}{e^{\phi}}
-\frac{\mathcal{R}f_{\mathcal{R}}}{2}+\frac{f}{2}\\\nonumber
&+&\left.f'_{\mathcal{R}}\left(\frac{\phi'}{2e^{\phi}}-\frac{\psi'}{2e^{\phi}}
-\frac{1}{re^{\phi}}\right)-\frac{f''_{\mathcal{R}}}{e^{\phi}}\right].
\end{eqnarray}
The quantity $\Omega$ in equation \eqref{g12} turns out to be
\begin{align}
\nonumber \Omega &=
\frac{2}{\left(\mathcal{R}f_{\mathcal{Q}}+2(1+f_{\mathcal{T}})\right)}\left[f'_{\mathcal{Q}}e^{-\phi}\mathcal{P}\left(\frac{1}{r^2}-\frac{e^\phi}{r^2}
+\frac{\psi'}{r}\right)+f_{\mathcal{Q}}e^{-\phi}\mathcal{P}\left(\frac{\psi''}{r}-\frac{\psi'}{r^2}-\frac{\phi'}{r^2}\right.\right.\\\nonumber
&-\left.\frac{\psi'\phi'}{r}-\frac{2}{r^3}+\frac{2e^\phi}{r^3}\right)+\mathcal{P}'\left\{f_{\mathcal{Q}}e^{-\phi}\left(\frac{\psi'\phi'}{8}
-\frac{\psi''}{8}-\frac{\psi'^2}{8}+\frac{\phi'}{2r}+\frac{\psi'}{2r}+\frac{1}{r^2}-\frac{e^{\phi}}{r^2}\right)\right.\\\nonumber
&+\left.\frac{3}{4}f_{\mathcal{T}}\right\}+\mathcal{P}f'_{\mathcal{T}}-\mu
f'_{\mathcal{T}}-\mu'\left\{\frac{f_{\mathcal{Q}}e^{-\phi}}{8}\left(\psi'^2-\psi'\phi'+2\psi''+\frac{4\psi'}{r}\right)
+\frac{3f_{\mathcal{T}}}{2}\right\}\\\nonumber
&-\left.\left(\frac{e^{-\phi}}{r^2}-\frac{1}{r^2}+\frac{\psi'e^{-\phi}}{r}\right)\left(\mu'f_{\mathcal{Q}}+\mu
f'_{\mathcal{Q}}\right)\right].
\end{align}

\end{document}